\documentclass[numberedappendix]{emulateapj}

\usepackage{amsmath}
\usepackage{color}
\usepackage{float}

\newcommand{\unitspace}{\ensuremath{\,}}
\newcommand{\usp}{\unitspace}

\newcommand{\unitstyle}[1]{\ensuremath{\mathrm{#1}}}
\newcommand{\power}[2]{\ensuremath{{#1}^{#2}}}

\newcommand{\kilo}{\unitstyle{k}}

\newcommand{\cm}{\unitstyle{cm}}
\newcommand{\gram}{\unitstyle{g}}

 \newcommand{\erg}{\unitstyle{erg}}

\newcommand{\cmsqperg}{\power{\cm}{2}\usp\power{\gram}{-1}}

   \newcommand{\eV}{\unitstyle{eV}}        \newcommand{\keV}{\kilo\eV}

\def\be{\begin{equation}}
\def\ee{\end{equation}}
\def\ba{\begin{eqnarray}}
\def\ea{\end{eqnarry}}
\def\bal#1\eal{\begin{align}#1\end{align}}

\shorttitle{Model atmospheres for X-ray bursting neutron stars}

\begin{document}
\title{Model atmospheres for X-ray bursting neutron stars}

\author{Zach Medin$^{1}$, Marina von Steinkirch$^{2,3}$, Alan C. Calder$^{2,3}$,\\ Christopher J. Fontes$^{1}$, Chris L. Fryer$^{1,4,5}$, and Aimee L. Hungerford$^{1}$}
\affil{$^1$Los Alamos National Laboratory, Los Alamos, NM 87545, USA\\
$^2$Department of Physics \& Astronomy, Stony Brook University, Stony Brook, NY 11794, USA\\
$^3$Institute for Advanced Computational Science, Stony Brook University, Stony Brook, NY 11794, USA\\
$^4$Physics Department, University of Arizona, Tucson, AZ 85721, USA\\
$^5$Physics and Astronomy Department, University of New Mexico, Albuquerque, NM 87131, USA}

\begin{abstract}
The hydrogen and helium accreted by X-ray bursting neutron stars is
periodically consumed in runaway thermonuclear reactions that cause
the entire surface to glow brightly in X-rays for a few seconds. With
models of the emission, the mass and radius of the neutron star can be
inferred from the observations. By simultaneously probing neutron star
masses and radii, X-ray bursts are one of the strongest diagnostics of
the nature of matter at extremely high densities. Accurate
determinations of these parameters are difficult, however, due to the
highly non-ideal nature of the atmospheres where X-ray bursts
occur. Observations from X-ray telescopes such as RXTE and NuStar can
potentially place strong constraints on nuclear matter once
uncertainties in atmosphere models have been reduced. Here we discuss
current progress on modeling atmospheres of X-ray bursting neutron
stars and some of the challenges still to be overcome.
\end{abstract}

\keywords{stars: neutron –-- X-rays: binaries --- X-rays: bursts}

\section{Introduction}
\label{sec:intro}

Neutron stars are ideal laboratories for the study of nuclear physics
in the extreme. In the core of neutron stars, conditions are such that
both exotic particles (e.g., hyperons, kaons, pions) and mixed phases
of hadronic and deconfined quark matter may exist. The high densities
and temperatures of these neutron stars can produce a broad suite of
matter states including crystalline, gapless superconducting, and
color-flavor-locked phases. All of this physics affects the equation
of state of nuclear matter, and ultimately places constraints on the
mass distribution, maximum neutron star mass, and the mass-radius
relation of neutron stars. But to study this physics, we must connect
the effects of the physics to observable features of neutron stars.

At present, there exist over 40 measured masses for neutron stars,
ranging from as low as 1.0M$_\odot$ to above 2.0M$_\odot$ \citep[for
  reviews, see][]{lattimerprakash2007,lattimer2012}. The maximum
stable neutron star mass can place constraints on the equation of
state and the behavior of matter at nuclear densities
(\citealt{glendenning1998,lattimerprakash2004,lackeyetal2006,
  lattimerprakash2007,schulzeetal2006, kurkelaetal2010}; see also
\citealt{demorestetal2010} for recent observational results). Coupled
mass/radius observations can place even stronger constraints on the
the equation of state for dense matter
\citep{ozel09,ozelbaymguver2010,steiner10,steiner13,miller2013}. While
X-ray bursts on neutron stars
\citep{vanparadijs1979,vanparadijs1986,ebisuzaki1987,damenetal1990,vanparadijsetal1990,madej04,majczynaetal2005,suleimanov11,spw11,miller11,spw12,gallowaylampe2012,poutanen14,nattila15,kajava16},
and thermal emission from quiescent and isolated neutron stars
\citep{rutledge.ea.99:refit,heinkeetal2006,hoetal2007,hoheinke2009,cackettetal2010,catuneanuetal2013,guillotetal2013,klochkov15,ofengeim15}
remain the most-studied observational probes providing simultaneous
measurements of mass and radius, a growing list of observations have
been proposed to provide this coupled data, from pulsar timing to
gravitational wave signals from merging neutron stars \citep[for a
  review, see][]{lattimer2012}.

During an X-ray burst (XRB), the accreted matter in the outer layers
of the neutron star undergoes thermonuclear burning, heating the
atmosphere and causing it to expand due to increased radiation
pressure. In the hottest bursts, the so-called ``photospheric radius
expansion'' (PRE) bursts, the heating is strong enough that the
atmosphere expansion is observed; in addition, the outgoing radiation
spectrum from the PRE source changes rapidly, both during expansion
and in the cooling/contraction at the end of the burst
\citep{vanparadijsetal1990,damenetal1990,lewin93}. Because of the
highly variable nature of these sources, observations at several time
points during a PRE burst permit tighter constraints on the mass and
radius than a single observation would. For example,
\citet{suleimanov11} constrained the mass and radius of 4U~1724--307
to a composition-dependent curve in mass-radius space by fitting
models of the PRE spectrum as a function of time\footnote{Or rather,
  Suleimanov et al.\ fit the PRE spectrum as a function of outgoing
  flux, which changes over the course of the burst.} to
observations. This constraint curve could be reduced to a point if
either the distance to the source or the outgoing flux at the burst
peak were known.

The strength of these constraints is limited by several important
uncertainties associated with XRBs. Two of the largest uncertainties
are in the distances to the XRB sources and their atmospheric
compositions \citep[e.g.,][]{strohmayerbildsten2006,lattimer14}. There
are also uncertainties in how the physics of the burst correlates with
the observations, for example, at what times during an observation the
photosphere is expanding, contracting, or has reached ``touchdown'';
i.e., at what time the photosphere has returned to (approximately) its
pre-burst radius \citep[cf.][]{ozel09,steiner10}. Finally, there are
uncertainties in the model approximations and techniques, which lead
to discrepancies in the results of different modeling
groups. Discrepancies are seen even between groups that include
similar physics in their models, such as \citet{madej04} and
\citet{spw12}: for these two groups, the temperature profiles in the
atmosphere and the outgoing spectra differ at the tens-of-percent
level and have qualitatively different shapes \citep[see appendix~C
  of][]{spw12}.

It is not surprising that there are discrepancies between the models,
considering the complexity of the physics involved in modeling
XRBs. At a minimum, these models require multi-frequency, multi-angle
radiation transfer with absorption and an exact Compton scattering
treatment, and hydrostatic balance with gas and radiation pressure
\citep[cf.][]{spw11,spw12}; ideally the models also include general
relativity and hydrodynamics. In addition, solving the equations can
be challenging numerically, due to the large differences in scale
across the atmosphere. For example, the gas pressure is several orders
of magnitude smaller than the radiation pressure in the outer
atmosphere.

In this paper we describe our efforts to reduce uncertainty in XRB
models, focusing specifically on the discrepancies seen between the
results of Madej et al.\ and Suleimanov et al. Like these two groups,
we model the outgoing spectra from neutron star atmospheres under a
suite of XRB conditions, using the full Boltzmann equation for Compton
scattering rather than the Fokker-Planck (Kompaneets)
approximation. However, we solve the equations of radiation transfer
in a time-dependent manner using an implicit Monte Carlo scheme
\citep{fleckcummings1971,canfieldetal1987}, rather than the
time-independent, deterministic methods of the other two groups. For
improved accuracy, we use detailed absorption opacities from the Los
Alamos National Laboratory (LANL) OPLIB database and include the
stimulated scattering contribution to the scattering opacity; for
completeness, we also consider general relativistic effects. Our
approach poses unique challenges, but also provides us independent
results with model uncertainties that are uncorrelated with those of
other XRB modeling groups.

The plan of this paper is as follows: In Section~\ref{sec:equations}
we discuss the physics involved in our XRB models and the equations to
be solved. In Section~\ref{sec:method} we discuss our model and
simulation methodology, including our treatment of Monte Carlo
radiation transport and hydrostatic balance. The dependence of our
results on the different pieces of physics is discussed in
Section~\ref{sec:physics}. These results are compared to the
calculations of both \citet{madej04} and \citet{spw12} in
Section~\ref{sec:results}. We conclude in Section~\ref{sec:discuss}
with a summary of this work and a discussion of future studies to
address further uncertainties.
 
\section{Equations of radiation transfer, material energy, and hydrostatic balance}
\label{sec:equations}

There are three equations we use to model the state of the neutron
star atmosphere. The radiation transfer equation
(Section~\ref{sub:rteq}) describes the movement of the radiation field
from the hot interior of the neutron star out to the surface, as it
exchanges energy with the atmosphere material through
absorption/emission and inelastic scattering. Similarly, the material
energy equation (Section~\ref{sub:meeq}) describes the
radiation-material energy exchange from the material side. Finally,
the hydrostatic balance equation (Section~\ref{sub:hydroeq}) describes
the density distribution of the atmosphere in the equilibrium
situation where there is no bulk radial motion of the material, due to
the balance between gravity and other external forces (e.g., pressure
gradients). As is discussed in Section~\ref{sub:radtrans}, we ignore
other forms of energy transport in our models. In addition, since we
restrict our simulation domain to layers of the neutron star well
above the hydrogen and helium burning layer
(Section~\ref{sec:method}), we also ignore energy generation due to
nuclear processes.

\subsection{Radiation transfer equation}
\label{sub:rteq}

In the frame comoving with the material, the \mbox{time-,}
\mbox{angle-}, and frequency-dependent equation of radiation transfer
is \citep[e.g.,][]{lindquist1966,mihalasmihalas1984}
\bal
\nu^3\frac{d\left[I_\nu(\hat{\Omega})\nu^{-3}\right]}{d\lambda} = -\rho\kappa_\nu^{\rm tot}(\hat{\Omega}) I_\nu(\hat{\Omega}) + j_\nu^{\rm tot}(\hat{\Omega}) \,,
\label{eq:RTE}
\eal
where $d/d\lambda$ is the ``directional'' (in phase space) derivative
along the geodesic defined by affine parameter $\lambda$, $I_\nu$ is
the radiation specific intensity,
\be
\kappa_\nu^{\rm tot}(\hat{\Omega})=\kappa_\nu+\kappa_\nu^{\rm sc}(\hat{\Omega})
\label{eq:knu}
\ee
is the total attenuation opacity with $\kappa_\nu$ the absorption
opacity and $\kappa_\nu^{\rm sc}$ the scattering opacity, and
\be
j_\nu^{\rm tot}(\hat{\Omega})=j_\nu+j_\nu^{\rm sc}(\hat{\Omega})
\label{eq:jnu}
\ee
is the total emission coefficient with $j_\nu$ the emission
coefficient and $j_\nu^{\rm sc}$ the scattering emission
coefficient. Here $\hat{\Omega}$ is the direction of propagation and
$\nu$ is the frequency; we did not explicitly state it, but the
intensity and the various coefficients depend on the time $t$ and the
position $\mathbf{r}$ as well. Note that (as is standard practice)
$j_\nu$ represents spontaneous emission only while the $k_\nu I_\nu$
term represents the difference between the absorption and stimulated
emission \citep[e.g.,][]{rybickilightman1986,castor2004}. We assume
spherical symmetry but not isotropy (Section~\ref{sub:radtrans}), such
that Equation~(\ref{eq:RTE}) becomes equation~3.7
of \citet{lindquist1966}; applying this latter equation to our neutron
star atmosphere model and using
\be
\frac{\partial \ln {\cal R}}{\partial r} = \frac{g_r{\cal V}}{c^2}
\label{eq:lnR}
\ee
we obtain
\bal
\frac{1}{c}\frac{\partial I_\nu(\hat{\Omega})}{{\cal R}\partial t} + \mu\frac{\partial I_\nu(\hat{\Omega})}{{\cal V}\partial r} + \frac{1-\mu^2}{r}\left(1-r\frac{g_r{\cal V}}{c^2}\right)\frac{\partial I_\nu(\hat{\Omega})}{{\cal V}\partial \mu} \nonumber\\
 - \mu\frac{g_r{\cal V}}{c^2} \nu^4\frac{\partial \left[I_\nu(\hat{\Omega})\nu^{-3}\right]}{{\cal V}\partial \nu} = -\rho\kappa_\nu^{\rm tot}(\hat{\Omega}) I_\nu(\hat{\Omega}) + j_\nu^{\rm tot}(\hat{\Omega}) \,,
\label{eq:GRTE}
\eal
where
\be
\mu = \hat{\Omega} \cdot \hat{r} \,,
\ee
\be
{\cal V} = \left(1-\frac{2GM}{c^2r}\right)^{-1/2}
\label{eq:calV}
\ee
is the volume correction factor,
\be
{\cal R} = \left(1-\frac{2GM}{c^2r}\right)^{1/2}
\label{eq:calR}
\ee
is the redshift correction factor, and
\be
g_r = \frac{GM}{r^2}{\cal V}
\label{eq:ggrav}
\ee
is the local gravitational acceleration (see \citealt{thorne1977} and
references therein). Here $M$ is the total (rest, energy, and
gravitational) mass of the neutron star, $G$ is the gravitational
constant, and $c$ is the speed of light. As is mentioned in
\citet{lindquist1966}, in Equation~(\ref{eq:GRTE}) the terms
proportional to $\partial \ln {\cal R}/\partial r = g_r{\cal V}/c^2$
are due to gravitational effects: the $\partial I_\nu/\partial \mu$
term represents the effect of gravitational light bending on the
radiation, while the $\partial (I_\nu\nu^{-3})/\partial \nu$ term
represents the effect of gravitational redshift. To derive
Equation~(\ref{eq:GRTE}) we have assumed that the mass of the
atmosphere is much less than $M$, that the atmosphere pressure $P$ is
much less than $Mc^2/4\pi r^3$, and that gravity has a much greater
effect on the radiation intensity than does any motion of the
material. These approximations are discussed in
Section~\ref{sub:gravity}. Note that under the conditions of our
model, ${\cal V} = {\cal R}^{-1} = 1+z$, where $z$ is the fractional
change in the photon wavelength due to redshift
\citep[e.g.,][]{madej04,spw11}; however, in the equations of this
section we choose to keep ${\cal V}$ and ${\cal R}$ as separate
factors in order to show clearly the contributions due to time-like
general relativistic effects versus space-like ones. Note also that we
have added a subscript ``$r$'' to the gravitational acceleration to
remind the reader that this quantity depends on radius (unlike in,
e.g., \citealt{spw11}, where $g$ is fixed).

We assume local thermodynamic equilibrium (LTE) throughout the
atmosphere (Section~\ref{sub:opacity}), such that the absorption
opacity and emission coefficient are related by Kirchoff's law of
thermal radiation:
\be
j_\nu = \rho\kappa_\nu B_\nu(T) \,;
\label{eq:kirchoff}
\ee
while $\kappa_\nu$ and the absorption-only (i.e., uncorrected for
stimulated emission) opacity $\bar{\kappa}_\nu$ are related by
\be
\kappa_\nu = \bar{\kappa}_\nu \left(1-e^{-h\nu/k_{\rm B}T}\right) \,.
\label{eq:kappa}
\ee
Here
\be
B_\nu(T) = \frac{2h\nu^3/c^2}{\exp(h\nu/k_{\rm B}T)-1}
\ee
is the frequency-dependent Planck function, $h$ is the Planck
constant, and $k_{\rm B}$ is the Boltzmann constant. We use the LANL
OPLIB database (Section~\ref{sub:opacity}) to find the absorption
opacity in Equations~(\ref{eq:knu}) and (\ref{eq:kirchoff}). This
opacity is a function of the rest mass density $\rho$ and the material
temperature $T$ (assuming a single material temperature for both the
electrons and ions; Section~\ref{sub:plasma}), and includes the
correction for stimulated emission $1-e^{-h\nu/k_{\rm B}T}$.

The scattering opacity including Compton scattering is given by
\bal
\kappa_\nu^{\rm sc} = {}& \frac{n_e}{\rho}\int_0^\infty d\nu' \int_{4\pi} d\Omega' \, \sigma^{\rm sc}(\nu \rightarrow \nu', \hat{\Omega} \cdot \hat{\Omega}') \nonumber\\
 {}& \qquad\qquad\qquad\qquad \times \left[1+\frac{c^2}{2h{\nu'}^3}I_{\nu'}(\hat{\Omega}')\right] \,,
\label{eq:kappascat}
\eal
where $n_e$ is the electron number density and $\sigma^{\rm sc}$ is
the double differential (differential with respect to both angle and
frequency) scattering cross section. In Equation~(\ref{eq:kappascat})
and the remainder of the scattering equations in this section,
`unprimed' variables represent the state of the particles (photons or
electrons) before scattering, while the `primed' variables represent
the state after scattering, e.g., $\nu$ versus $\nu'$. We assume that
$n_e$ for the Compton scattering opacity represents the number density
of all electrons, bound or free; see
Section~\ref{sub:opacity}. Therefore, when using
Equation~(\ref{eq:kappascat}) we make the substitution
\be
n_e \simeq \frac{Y_e \rho}{m_u} \,,
\label{eq:ne}
\ee
where $Y_e = \langle Z \rangle/\langle A \rangle$ is the electron
fraction, $\langle Z \rangle$ and $\langle A \rangle$ are the
ion-averaged charge and atomic mass of the atmosphere mixture, and
$m_u$ is the atomic mass unit. Note that Equation~(\ref{eq:ne}) is an
approximation in the sense that the average nucleon mass of the
accreted material is not exactly $m_u$. The
$1+c^2I_{\nu'}(\hat{\Omega}')/2h{\nu'}^3$ factor in
Equation~(\ref{eq:kappascat}) is the correction due to stimulated
scattering, a quantum-mechanical phenomenon analogous to stimulated
emission where photons are more likely to be scattered into a densely
populated state \citep[e.g.,][]{wienke12}. Stimulated scattering is
important at the high temperatures and densities considered in this
paper (Section~\ref{sub:opacity}). For a Maxwell distribution of
electrons $f(p)$ with
\be
\int d\mathbf{p} \, f(p) = 1
\ee
we have \cite[e.g.,][]{wienke85,spw12}
\bal
\sigma^{\rm sc}(\nu \rightarrow \nu', \hat{\Omega} \cdot \hat{\Omega}') = {}& \frac{1}{2\pi} \int \frac{d\mathbf{p}}{\gamma} xx' f(p) \sigma_{\rm KN}(\nu_0 \rightarrow \nu'_0) \nonumber\\
 {}& \times \delta\left(xx'(1-\eta) - (x_0' - x_0)\right) \,,
\label{eq:sigmascat}
\eal
where $\delta()$ is the Dirac delta function and
\bal
\sigma_{\rm KN}(\nu_0 \rightarrow \nu'_0) = {}& \frac{3\sigma_{\rm Th}}{8x_0^2} \left[\frac{x_0}{x'_0} + \frac{x'_0}{x_0} \right. \nonumber\\
 {}& \left. + 2\left(\frac{1}{x_0}-\frac{1}{x'_0}\right) + \left(\frac{1}{x_0}-\frac{1}{x'_0}\right)^2\right]
\eal
is the Klein-Nishina differential cross section with
\be
x = \frac{h\nu}{m_ec^2} \,, \qquad\qquad x'= \frac{h\nu'}{m_ec^2} \,,
\ee
etc. Here $\mathbf{p} = p\hat{n}$ is the electron momentum with
\be
p = \gamma\beta m_ec^2
\label{eq:p}
\ee
and $\hat{n}$ the direction of propagation,
\be
\gamma = \frac{1}{\sqrt{1-\beta^2}}
\ee
is the electron Lorentz factor, $\beta = v_e/c$ is the ratio of the
electron velocity to the speed of light, and
\be
\nu_0 = \gamma \nu(1-\beta\zeta)
\label{eq:nu0}
\ee
and
\be
\nu_0' = \gamma \nu'(1-\beta\zeta')
\label{eq:nu0p}
\ee
are the photon frequencies in the rest frame of the pre-scattered
electron; note that the electron quantities in the above equations
($p$, $\beta$, etc.) are all at their pre-scattering values. In
addition,
\be
\eta = \hat{\Omega} \cdot \hat{\Omega}' \,,
\ee
\be
\zeta = \hat{\Omega} \cdot \hat{n} \,,
\ee
\be
\zeta' = \hat{\Omega}' \cdot \hat{n} \,,
\ee
$\sigma_{\rm Th}$ is the Thomson cross section, and $m_e$ is the
electron rest mass. For simplicity, in our calculations we use the
(non-relativistic) Maxwell-Boltzmann electron distribution function,
given by
\be
f(p) = \left(\frac{m_ec^2}{2\pi k_{\rm B}T}\right)^{3/2} e^{-p^2/2m_ec^2k_{\rm B}T}
\ee
\citep[but see][]{canfieldetal1987}. After orienting the direction
variables relative to $\hat{\Omega}$, with $\hat{\Omega}$ acting as
the $\hat{z}$ axis, Equation~(\ref{eq:kappascat}) becomes
\bal
\kappa_\nu^{\rm sc} = {}& \frac{Y_e}{2\pi m_u} \int_0^\infty d\nu' \int_{-1}^1 d\eta \int_0^{2\pi} d\eta_\perp \nonumber\\
 {}& \times \int_0^\infty dp \int_{-1}^1 d\zeta \int_0^{2\pi} d\zeta_\perp \frac{xx'}{\gamma} p^2 f(p) \sigma_{\rm KN}(\nu_0 \rightarrow \nu'_0) \nonumber\\
 {}& \times \left[1+\frac{c^2}{2h{\nu'}^3}I_{\nu'}(\hat{\Omega}')\right] \delta\left(xx'(1-\eta) - (x_0' - x_0)\right) \,;
\eal
here $\eta_\perp$ and $\zeta_\perp$ are the azimuthal angles with
respect to the polar angles $\eta$ and $\zeta$, respectively. Changing
variables using Equations~(\ref{eq:nu0}) and (\ref{eq:nu0p}),
\be
\int d\eta \int d\eta_\perp = \frac{\nu_0 \nu'_0}{\nu \nu'} \int d\eta_0 \int d\eta_{0,\perp} \,,
\ee
\be
\int d\nu' = \frac{\nu'}{\nu'_0} \int d\nu'_0 \,,
\ee
and the invariant
\be
\nu\nu'(1-\eta) = \nu_0\nu'_0(1-\eta_0)
\ee
gives
\bal
\kappa_\nu^{\rm sc} = {}& \frac{Y_e}{m_u} \int_0^\infty 4\pi p^2 dp \, f(p) \int_{-1}^1 d\zeta \frac{1-\beta\zeta}{2} \nonumber\\
 {}& \times \int_0^{2\pi} \frac{d\zeta_\perp}{2\pi} \int_0^{2\pi} \frac{d\eta_{0,\perp}}{2\pi} \int_{\nu_0/(1+2x_0)}^{\nu_0} d\nu'_0 \nonumber\\
 {}& \times \sigma_{\rm KN}(\nu_0 \rightarrow \nu'_0) \left[1+\frac{c^2}{2h{\nu'}^3}I_{\nu'}(\hat{\Omega}')\right] \,.
\label{eq:kappascat2}
\eal
We have expressed the scattering opacity in the above form for ease of
sampling using the Monte Carlo method (see the Appendix).

The scattering emission coefficient (including Compton scattering and
the correction for stimulated scattering) is given by
\bal
j_\nu^{\rm sc} = {}& \frac{Y_e \rho}{m_u} \left[1+\frac{c^2}{2h{\nu}^3}I_\nu(\hat{\Omega})\right] \nonumber\\
 {}& \times \int_0^\infty d\nu' \int_{4\pi} d\Omega' \frac{\nu}{\nu'} \sigma^{\rm sc}(\nu' \rightarrow \nu, \hat{\Omega} \cdot \hat{\Omega}') I_{\nu'}(\hat{\Omega'})
\label{eq:jscat}
\eal
[cf.\ Equation~(\ref{eq:kappascat})].

Integrating both sides of Equation~(\ref{eq:GRTE}) over angle and
frequency (using integration by parts with $[I_\nu]_{\nu=0}^\infty =
0$, $[I_\nu]_{\phi_\Omega=0}^{2\pi} = 0$, etc.) and using
Equation~(\ref{eq:lnR}) gives
\bal
\frac{\partial u}{{\cal R}\partial t} + \frac{1}{r^2{\cal R}^2}\frac{\partial \left(r^2F{\cal R}^2\right)}{{\cal V}\partial r} \nonumber\\
 = \int_0^\infty d\nu \int_{4\pi} d\Omega \left[-\rho\kappa_\nu^{\rm tot}(\hat{\Omega}) I_\nu(\hat{\Omega}) + j_\nu^{\rm tot}(\hat{\Omega})\right] \,,
\label{eq:F0thGRTE}
\eal
where
\be
u = \frac{1}{c}\int_0^\infty d\nu \int_{4\pi} d\Omega \, I_\nu(\hat{\Omega})
\ee
is the total radiation energy density and
\be
F = \int_0^\infty d\nu F_\nu = \int_0^\infty d\nu \int_{4\pi} d\Omega \, \mu I_\nu(\hat{\Omega})
\label{eq:flux}
\ee
is the total radiation flux in the outward radial direction. As is
mentioned in \citet{thorne1967}, one of the factors of ${\cal R}$ in
the $\partial (r^2F{\cal R}^2)/\partial r$ term of
Equation~(\ref{eq:F0thGRTE}) represents the gravitational redshift of
the transported radiation, while the other represents the time
dilation. Note that exterior to the neutron star, in steady state
Equation~(\ref{eq:F0thGRTE}) gives
\be
\frac{\partial \left(L_r{\cal R}^2\right)}{\partial r} = 0
\ee
or
\be
L_r{\cal R}^2 \equiv \frac{L_r}{(1+z)^2} = L_\infty \,,
\label{eq:luminosity}
\ee
where $L_r = 4\pi r^2F$ is the neutron star luminosity as seen by an
observer at distance $r$ from the center of the star and $L_\infty$ is
the luminosity as seen by an observer at infinity (Earth).

Similarly, taking the first angular moment of Equation~(\ref{eq:GRTE})
and integrating over frequency gives
\bal
\frac{1}{c^2}\frac{\partial F}{{\cal R}\partial t} + \frac{1}{{\cal V}}\left(\nabla \cdot \mathbf{P}\right)_r + \rho g_r\left(\frac{u+P_{rr}}{\rho c^2}\right) \nonumber\\
 = \frac{1}{c} \int_0^\infty d\nu \int_{4\pi} d\Omega \, \mu \left[-\rho\kappa_\nu^{\rm tot}(\hat{\Omega}) I_\nu(\hat{\Omega}) + j_\nu^{\rm tot}(\hat{\Omega})\right] \,,
\label{eq:FA1stGRTE}
\eal
where
\be
P_{rr} = \frac{1}{c} \int_0^\infty d\nu \int_{4\pi} d\Omega \, \mu^2 I_\nu(\hat{\Omega})
\ee
is the $rr$ component of the total radiation pressure tensor
$\mathbf{P}$; here
\be
\left(\nabla \cdot \mathbf{P}\right)_r = \frac{1}{r^2}\frac{\partial \left[r^2P_{rr}\right]}{\partial r} + \frac{P_{rr}-u}{r}
\label{eq:Ptens}
\ee
\citep[e.g.,][]{castor2004}. The $u+P_{rr}$ term in
Equation~(\ref{eq:FA1stGRTE}) represents the gravitational attraction
of the photon gas toward the center of the star \citep{thorne1967}.

\subsection{Material energy equation}
\label{sub:meeq}

We assume infinite ion-electron coupling and ignore heat conduction
(the validity of these assumptions is discussed in
Section~\ref{sec:physics}), such that the material energy equation is
\be
\rho c_V\frac{\partial T}{{\cal R}\partial t} = \int_0^\infty d\nu \int_{4\pi} d\Omega \left[\rho\kappa_\nu^{\rm tot}(\hat{\Omega}) I_\nu(\hat{\Omega}) - j_\nu^{\rm tot}(\hat{\Omega})\right] \,,
\label{eq:MEE}
\ee
where $c_V$ is the total (ion plus electron) specific heat. We also
assume an ideal gas equation of state, such that
\be
c_V = \frac{3k_{\rm B}}{2m_u}\frac{1+\langle Z \rangle}{\langle A \rangle}
\ee
and
\be
P_{\rm gas} = \frac{\rho k_{\rm B}T}{m_u}\frac{1+\langle Z \rangle}{\langle A \rangle}
\label{eq:Pgas}
\ee
where $P_{\rm gas}$ is the gas pressure.

\subsection{Hydrostatic balance equation}
\label{sub:hydroeq}

The gravitational force per unit volume on the gas is
\be
\mathbf{f}_{\rm grav} = -\rho g_r \hat{r} \,,
\label{eq:fgrav}
\ee
where we have assumed that the gas pressure and internal energy
density are much smaller than the gas rest energy density $\rho c^2$
(see Section~\ref{sub:gravity}). The buoyancy force per unit volume on
the gas is [cf.\ Equation~(\ref{eq:FA1stGRTE})]
\bal
 {}& \mathbf{f}_{\rm buoy} = -\left\{\frac{\partial P_{\rm gas}}{{\cal V}\partial r} \right. \nonumber\\
 {}& \left. + \frac{1}{c} \int_0^\infty d\nu \int_{4\pi} d\Omega \, \mu \left[-\rho\kappa_\nu^{\rm tot}(\hat{\Omega}) I_\nu(\hat{\Omega}) + j_\nu^{\rm tot}(\hat{\Omega})\right] \right\} \hat{r} \,.
\label{eq:fbuoy}
\eal
We assume that the rotational force on the gas is much smaller than
the gravitational and buoyancy forces (Section~\ref{sub:hydro});
setting $\mathbf{f}_{\rm grav} = - \mathbf{f}_{\rm buoy}$
[Equations~(\ref{eq:fgrav}) and (\ref{eq:fbuoy})], we obtain our
equation of hydrostatic balance:
\bal
\frac{\partial P_{\rm gas}}{{\cal V}\partial r} = {}& -\frac{1}{c} \int_0^\infty d\nu \int_{4\pi} d\Omega \, \mu \left[-\rho\kappa_\nu^{\rm tot}(\hat{\Omega}) I_\nu(\hat{\Omega}) + j_\nu^{\rm tot}(\hat{\Omega})\right] \nonumber\\
 {}& \qquad\qquad\qquad\qquad - \rho g_r
\label{eq:hydroeq}
\eal
\citep[cf.][]{mihalasmihalas1984}. Note that if $P_{rr} = u/3$ (i.e.,
assuming the Eddington approximation), Equation~(\ref{eq:Ptens})
becomes
\be
\left(\nabla \cdot \mathbf{P}\right)_r = \frac{\partial P_{rr}}{\partial r} \,;
\label{eq:Ptens2}
\ee
in steady state, combining Equations~(\ref{eq:FA1stGRTE}),
(\ref{eq:hydroeq}), and (\ref{eq:Ptens2}), we recover the
Tolman-Oppenheimer-Volkoff equation \citep[e.g.,][]{thorne1977}:
\be
\frac{\partial P}{\partial r} \equiv \frac{\partial P_{\rm gas}}{\partial r} + \frac{\partial P_{rr}}{\partial r} = -\rho g_r{\cal V}\left(1+\frac{u+P_{rr}}{\rho c^2}\right) \,.
\label{eq:dPdr}
\ee
Alternatively, if $\partial P_{\rm gas}/\partial r < 0$ in the
atmosphere (i.e., there is no density inversion) and $\kappa_\nu^{\rm
  tot}$ and $j_\nu^{\rm tot}$ are isotropic, then
Equation~(\ref{eq:hydroeq}) requires
\be
F < F_{\rm crit}
\label{eq:Fcritcond}
\ee
with
\be
F_{\rm crit} = \frac{cg_r}{\kappa_F^{\rm tot}}
\label{eq:Fcrit}
\ee
and
\be
\kappa_F^{\rm tot} = \frac{\int_0^\infty d\nu \, \kappa_\nu^{\rm tot} F_\nu}{F}
\label{eq:kappaF}
\ee
for hydrostatic balance. Note that the critical ``luminosity'' for
this case, $4\pi r_{\rm surf}^2 F_{\rm crit,surf}$, is equivalent to
the Eddington luminosity
\be
L_{\rm Edd} = \frac{4\pi cGM}{\kappa_F^{\rm tot}} {\cal V}
\label{eq:LEdd}
\ee
\citep[cf.\ equations~2 and 5 of][]{spw12}.
 
\section{Model and simulation methodology}
\label{sec:method}

\subsection{General model}
\label{sub:model}

For a given neutron star atmosphere model, we solve for the
equilibrium structure and outgoing radiation spectrum in three steps:
we first fix the conditions at the base of the atmosphere, as
described below; then guess the initial density, temperature, and
radiation intensity in the atmosphere, as described in
Section~\ref{sub:initial}; and finally evolve the atmosphere to a
steady state, using the equations of Section~\ref{sec:equations}. In
each time step of the simulation, the radiation transfer and material
energy equations are solved using an implicit Monte Carlo method
\citep{fleckcummings1971}; see Section~\ref{sub:MC}. After several
time steps hydrostatic balance is restored by adjusting the atmosphere
density with Equation~(\ref{eq:hydroeq}); see
Section~\ref{sub:hydrosolve}. The equilibrium, outgoing radiation
spectrum obtained from the above procedure is then fit to a curve
defined by a color correction factor, as described in
Section~\ref{sec:results}; the fitted color correction factor can be
used to compare our model results with observations of X-ray bursts
\citep[as is done in, e.g.,][]{spw11,spw12}. Note that in this paper,
the term ``atmosphere'' effectively means ``the domain of our
simulation''. As is discussed in Section~\ref{sub:initial}, we choose
our simulation domain, and therefore our definition of atmosphere, to
extend to optical depths of around 100; this corresponds to densities
of a few~${\rm g~cm^{-3}}$.

For simplicity we assume that $\{X\}$, the chemical composition in the
atmosphere, is uniform and constant in time
(Section~\ref{sub:mixing}). We also assume that the radiation at the
base of the atmosphere is in thermal equilibrium with the material,
such that
\be
I_\nu(\hat{\Omega},r_{\rm base}) = B_\nu(T_{\rm base})
\label{eq:Ibase}
\ee
(Section~\ref{sub:radmat}), where $r_{\rm base}$ and $T_{\rm base}$
are the radius and temperature at the base of the atmosphere. With
these assumptions, in equilibrium our models are fully determined by
$\{X\}$, $M$, $r_{\rm base}$, and $T_{\rm base}$. However, in this
paper we use the alternate parameter set $\{X\}$, $r_{\rm base}$, the
gravitational acceleration at the base of the atmosphere $g_{\rm
  base}$, and the luminosity ratio
\be
l_{\rm proj} = \frac{L_{\rm proj}}{L_{\rm Th}};
\label{eq:lproj}
\ee
here 
\be
L_{\rm proj} = L_{\rm surf}\left(\frac{{\cal R}_{\rm surf}}{{\cal R}_{\rm base}}\right)^2
\label{eq:Lproj}
\ee
is the luminosity ``projected'' on to the base of the
atmosphere [Equation~(\ref{eq:luminosity})], $L_{\rm surf}$ is the
luminosity as seen by an observer at the surface, and
\be
L_{\rm Th} = \frac{4\pi cGM}{\kappa_{\rm Th}} {\cal V}_{\rm base}
\label{eq:LTh}
\ee
is the ``Thomson'' Eddington luminosity with
\be
\kappa_{\rm Th} = \frac{Y_e\sigma_{\rm Th}}{m_u}
\label{eq:kth}
\ee
[cf.\ Equation~(\ref{eq:LEdd})]. This parameter set, which follows
\citet{spw11,spw12}, also fully determines the equilibrium
atmosphere. However, it has two important advantages over the former
set: first, in thin atmospheres the equilibrium solution depends only
on $\{X\}$, $g_{\rm base}$, and $l_{\rm proj}$ (not $r_{\rm base}$);
second, unlike temperature, the luminosity can be tied directly to
observations of X-ray burst fluxes \citep[see below and][]{spw11}. The
main disadvantage of this set is that $M$ and $T_{\rm base}$ are
derived from our specified parameters. The neutron star mass $M$ is
easily found using Equation~(\ref{eq:ggrav}), but $T_{\rm base}$ must
be found using a shooting method or some similar technique: we guess a
value for $T_{\rm base}$, evolve the system using the equations of
Section~\ref{sec:equations} to obtain a steady-state value for $L_{\rm
  surf}$ and hence $L_{\rm proj}$, and then adjust $T_{\rm base}$ and
iterate until Equation~(\ref{eq:lproj}) is satisfied (i.e., until
$L_{\rm proj} = l_{\rm proj} L_{\rm Th}$). Note that using $L_{\rm
  proj}$ above rather than $L_{\rm surf}$ is desirable because the
former corresponds to a fixed location in space and therefore has less
of the position-related ambiguity associated with general relativistic
quantities. In addition, $L_{\rm proj}$ is related to the observed
flux $F_\infty$ by a constant ($F_\infty = L_{\rm proj} {\cal R}_{\rm
  base}^2/4\pi D^2$, where $D$ is the distance to Earth), and is
therefore more easily fit to observations
\citep[see][]{spw11,spw12}. In thin atmospheres $L_{\rm base} \simeq
L_{\rm surf}$, but in extended atmospheres the two luminosity measures
can be very different.

We keep $r_{\rm base}$ fixed at $11.5$~km, as is discussed in
Section~\ref{sub:gravity}; but we vary the other three parameters in a
manner similar to \citet{spw11,spw12} to obtain more than 100
different atmosphere models (see Section~\ref{sec:results}). The
material quantities in each model (e.g., temperature, density, and
opacity) are defined on a grid of 100 cells discretized in radius,
representing spherical shells of the atmosphere; because of the
assumed spherical symmetry of the model (Sections~\ref{sec:equations}
and \ref{sub:radtrans}), the other two coordinates (polar angle and
azimuthal angle) do not need to be specified. The cells extend from an
optical depth much greater than unity at the base of the atmosphere to
an optical depth much less than unity at the surface
(Section~\ref{sub:initial}). The centers of the cells are
approximately equally spaced in $r$; the spacing is chosen to be small
enough that the material quantities do not change too rapidly from
cell to cell, but large enough that there are sufficient Monte Carlo
particles in each cell at every time step (Section~\ref{sub:MC}). The
opacity tables used in the simulations are limited to a relatively low
number of frequency groups for faster data lookup and more manageable
storage; for convenience, the same set of frequency groups is used to
obtain color correction factors and generate the spectra plots in this
paper. Specifically, we use 300 logarithmically spaced groups in the
range $\nu = 10$~eV to $1$~MeV. The number of groups chosen is large
enough that the fitted color correction factors are converged in
frequency space, but small enough that there are several Monte Carlo
particles contributing to the outgoing spectrum for each group near
the spectral peak; while the range of groups chosen is large enough to
cover the vast majority of photons emitted and scattered in all of the
atmospheres considered here. The number and range of cells and
frequency groups used in our simulations are comparable to those of
previous works \citep[e.g.,][]{madej04,spw11}.

To account for energy transfer to and from the radiation field, the
material energies in each cell are updated at the end of every time
step of the simulation. Because the radiation field is represented by
stochastic particles (Section~\ref{sub:MC}), sometimes the energy
transferred can be larger than the energy already in the material,
which can cause numerical stability and energy conservation
problems. This is particularly true in the outer layers of the
atmosphere where the density is low and in hot atmospheres where
Compton scattering is important (Section~\ref{sub:radmat}). We
mitigate this effect in two ways: first, we do not allow any one
particular Monte Carlo particle to transfer more than 50\% of the
cell's energy to or from the cell (the remaining energy stays in the
Monte Carlo particle or is re-radiated as a new particle); and second,
we take sufficiently small time steps that the cell energy changes by
less than 50\% per time step over the majority of the cells and the
majority of the time steps in the calculation. Though the exact value
depends on the particular atmosphere model, we find that using $\Delta
t_{\rm surf} \simeq 10^{-9}$~s, where $\Delta t_{\rm surf}$ is the
time step in the outer cell of the simulation (see
Section~\ref{sub:MC}), is reasonable to prevent cell energy
problems. Note that for simplicity we do not use an adaptive time step
in our calculations.

\subsection{Initial conditions}
\label{sub:initial}

Ideally, using the above method we can find the equilibrium
atmosphere solution for any guess of the initial density, temperature,
and radiation intensity. However, we would like to start with a guess
that is close to the final solution, both to avoid cases that are
unstable to our method and to limit the number of iterations necessary
to reach a solution. Here we describe our procedure for obtaining such
an initial guess.

To obtain an initial guess for ``thin'' atmospheres (see below), we
assume that $F$, $g_r$, ${\cal V}$, and ${\cal R}$ are uniform in
space, and that the latter three are given by their values at the base
of the atmosphere while
\be
F = \frac{L_{\rm proj}}{4\pi r_{\rm base}^2} \,.
\label{eq:Fapprox}
\ee
In steady state, if $P_{rr} = u/3$, $\kappa_\nu^{\rm tot}$ and
$j_\nu^{\rm tot}$ are isotropic, and $u + P_{rr} \ll \rho c^2$ (none
of which are quite true; see Section~\ref{sub:gravity}), we have from
Equation~(\ref{eq:FA1stGRTE}) that
\be
\frac{1}{3}\frac{\partial u}{\partial y} = \frac{\kappa_F^{\rm tot}F}{c}
\label{eq:dudr}
\ee
and from Equation~(\ref{eq:hydroeq}) that
\be
\frac{\partial P_{\rm gas}}{\partial y} = g_{\rm base} - \frac{\kappa_F^{\rm tot}F}{c} \,.
\label{eq:hydroeq2}
\ee
Here $y$ is the column depth, given for thin atmospheres by
\be
\frac{\partial y}{{\cal V_{\rm base}}\partial r} = -\rho
\label{eq:dydr}
\ee
and
\be
y_{\rm surf} = 0 \,;
\label{eq:ysurf}
\ee
for the opacity in Equations~(\ref{eq:dudr}) and (\ref{eq:hydroeq2})
we use the expression of \citet{spw12},
\be
\kappa_F^{\rm tot} \simeq \kappa_{\rm Th} \left[1+\left(\frac{k_{\rm B}T}{38.8~\keV}\right)^{\alpha_g}\right]^{-1}
\label{eq:kpapprox}
\ee
with
\be
\alpha_g = 1.01+0.067\left(\log g_r - 14.0\right)
\ee
\citep[see also equation~2
  of][]{paczynski86}. Equation~(\ref{eq:kpapprox}) comes from
averaging the (Klein-Nishina) scattering opacity, which dominates the
total opacity in hot neutron star atmospheres, over a Maxwell
distribution of electrons and using the diffusion approximation for
the scattered radiation intensity [e.g., \citealt{sampson:opacity};
  cf.\ Equation~(\ref{eq:kappascat2})]. We also assume the
Milne-Eddington boundary condition
\be
u_{\rm surf} = \frac{2F}{c} \,,
\label{eq:milne}
\ee
and that the radiation field is given by a Planck function at a
radiation temperature $T_r$, such that
\be
u_\nu = \frac{4\pi}{c} B_\nu(T_r)
\label{eq:unu}
\ee
and
\be
u = \frac{4\pi}{c}\int_0^\infty d\nu B_\nu(T_r) \equiv aT_r^4
\label{eq:u}
\ee
with $a = 4\sigma_{\rm SB}/c$ the radiation constant and $\sigma_{\rm
  SB}$ the Stefan-Boltzmann constant. Note that in this section and
later analysis it is useful to use Equation~(\ref{eq:u}) more
generally, such that the radiation temperature is defined by $aT_r^4 =
u$ even in cases where the radiation field is not a Planck
function. In the inner layers of the neutron star atmosphere, $T
\simeq T_r$ (which does not mean that the material and the radiation
field have the same energy density there, only that they are in
thermal equilibrium with each other;
cf.\ Section~\ref{sub:opacity}). However, in the outer layers these
two quantities diverge due mainly to Compton downscattering of the
photons (see Section~\ref{sub:radmat}). Therefore, for simplicity we
assume that
\be
T =
\left\{
\begin{array}{ll}
T_r \,, & \rm{inner~layers}; \\
T_{\rm outer} \,, & \rm{outer~layers}
\end{array}
\right.
\label{eq:TTr}
\ee
where $T_{\rm outer}$ is constant with radius and the separation
between ``inner layers'' and ``outer layers'' is found by making $T$
continuous. We find the approximate temperature in the outer layers
$T_{\rm outer}$, along with the critical luminosity ratio at the
surface
\be
l_{\rm crit,surf} = \frac{L_{\rm surf}}{L_{\rm Edd}}
\label{eq:lcrit}
\ee
[cf.\ Equation~(\ref{eq:lproj})], by iteration: We first guess $T_{\rm
  outer} = T_{r,\rm surf}$, then update $l_{\rm crit,surf}$ using
Equations~(\ref{eq:LEdd}) and (\ref{eq:kpapprox}) with $T=T_{\rm
  outer}$, and finally recalculate $T_{\rm outer}$ using
\bal
T_{\rm outer} = {}& \left(\frac{F}{\sigma_{\rm SB}}\right)^{1/4}l^{3/20}\left(\frac{3+5X_{\rm H}}{1-l}\right)^{2/15} \nonumber\\
 {}& \qquad \times \left[(0.102+0.008X_{\rm H})\ln\frac{3+5X_{\rm H}}{1-l}\right. \nonumber\\
 {}& \qquad\qquad \left.+ 0.63-0.06X_{\rm H}\right]^{-4/5} \,,
\label{eq:Touter}
\eal
with $l=l_{\rm crit,surf}$; this process is repeated several times to
convergence. Here $X_{\rm H}$ is the mass fraction of hydrogen, and
Equation~(\ref{eq:Touter}) is from \citet{spw12} \citep[see
  also][]{londonetal1986,pavlovshibanovzavlin1991}. With these
assumptions, we can use Equations~(\ref{eq:Fapprox}), (\ref{eq:dudr}),
and (\ref{eq:ysurf})--(\ref{eq:milne}) to solve for $T_r$ as a
function of $y$; then Equation~(\ref{eq:TTr}) to solve for $T$ as a
function of $y$; then Equations~(\ref{eq:Pgas}) and
(\ref{eq:hydroeq2}) and the condition $\rho_{\rm surf} = 0$ to solve
for $\rho$ as a function of $y$; and finally Equation~(\ref{eq:dydr})
to solve for $y$ as a function of $r$. We choose the extent of $r$ in
our simulation such that the optical depths $\tau_F^{\rm tot}$ are
around 100 for the deepest cell and $10^{-6}$ for the shallowest cell,
using
\be
\frac{\partial \tau_F^{\rm tot}}{\partial y} = \kappa_F^{\rm tot} \,.
\label{eq:dtaudr}
\ee
Note that the minimum optical depth we use is comparable to the one
used in \citet{spw11,spw12}, but that the maximum optical depth is two
orders of magnitude smaller than the one used in that work. The limit
$\tau_F^{\rm tot} \alt 100$ is imposed by the Monte Carlo method we
use; see Section~\ref{sub:radmat}.

\begin{figure*}
\begin{center}
\begin{tabular}{c c}
\includegraphics[width=\columnwidth]{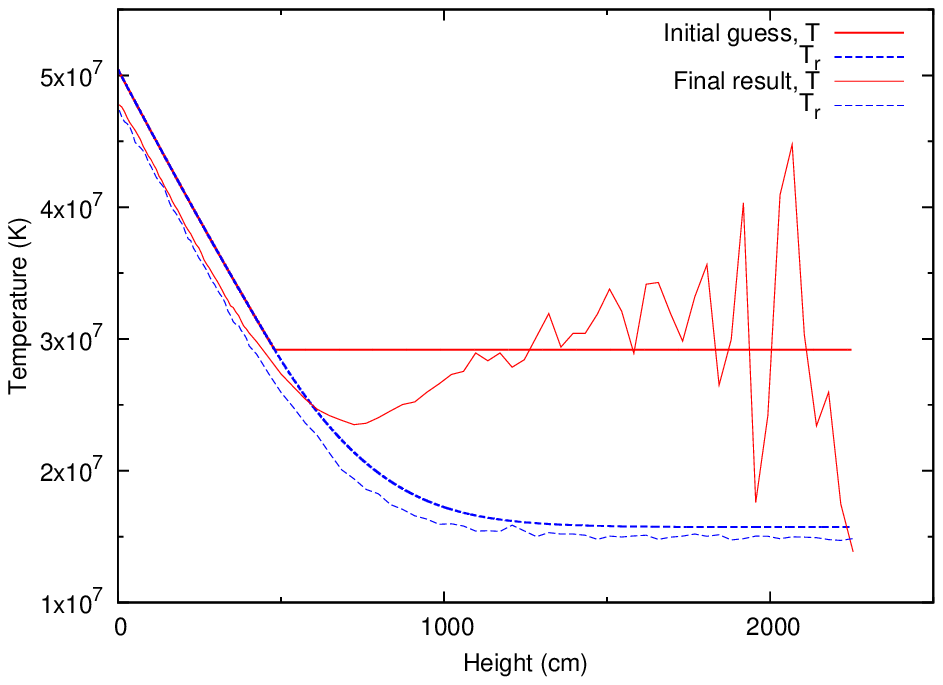} &
\includegraphics[width=\columnwidth]{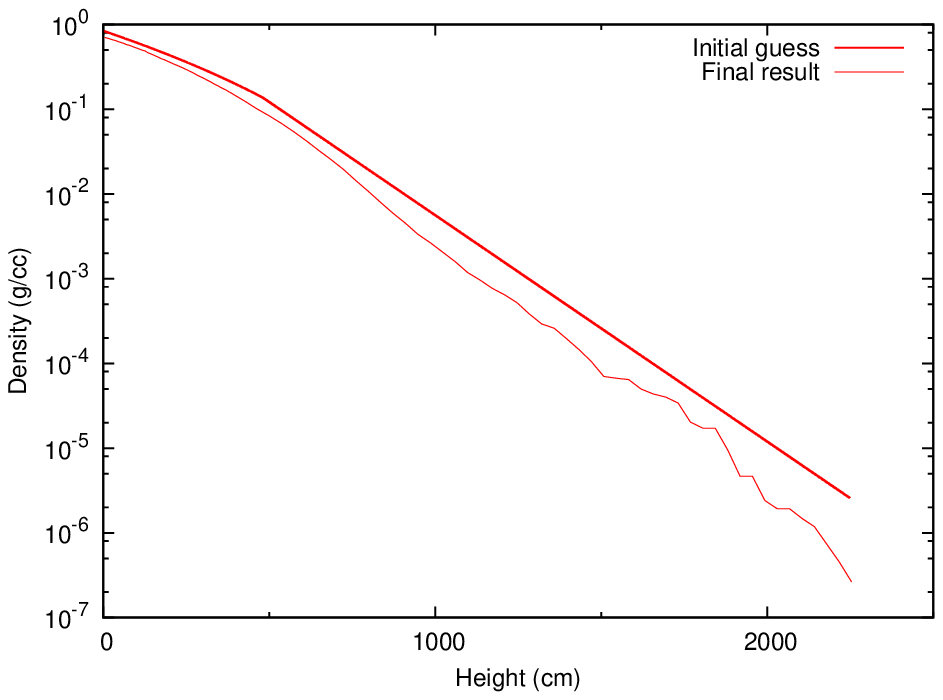} \\
\includegraphics[width=\columnwidth]{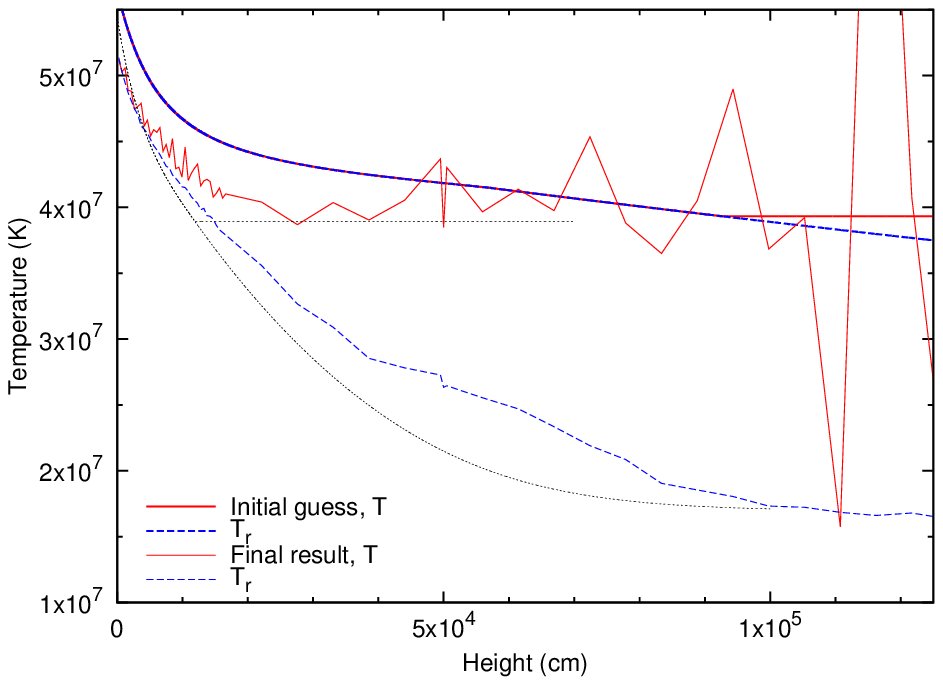} &
\includegraphics[width=\columnwidth]{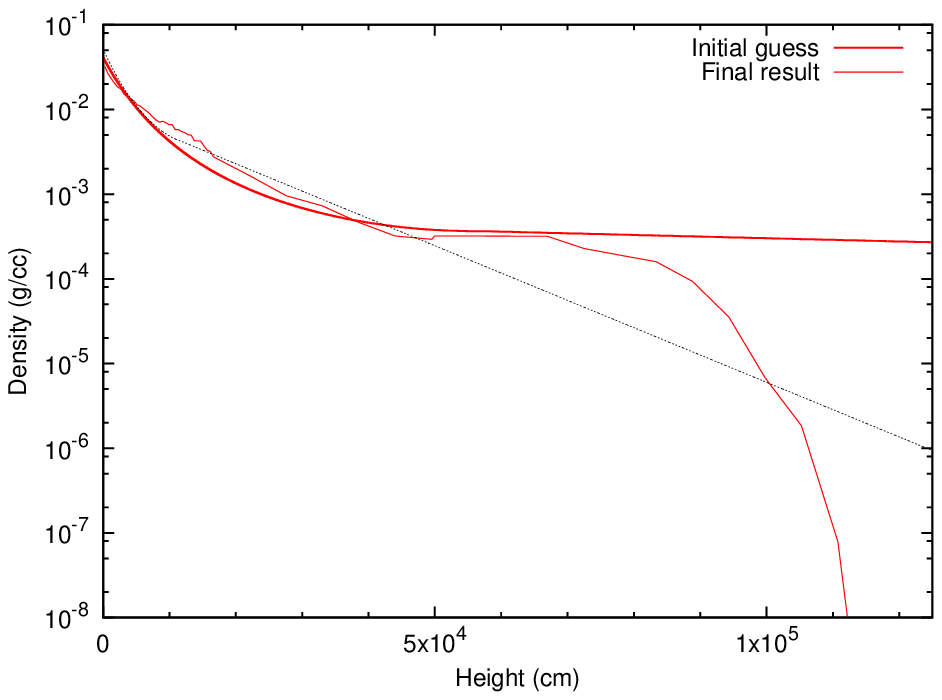}
\end{tabular}
\caption{Initial guess for the material and radiation temperature (the
  curves labeled ``$T$'' and ``$T_r$'', respectively; left panels) and
  the density (right panels) as a function of the radial distance
  above the atmosphere base $r-r_{\rm base}$, for atmosphere models
  with solar composition, surface gravity $g_{\rm base} = 10^{14}~{\rm
    cm~s^{-2}}$, and luminosity ratio $l_{\rm proj} = 0.8$ (top
  panels) or $l_{\rm proj} = 1.09$ (bottom panels). The temperature
  and density profiles at the end of the calculation are also shown
  (``Final result''; cf.\ Section~\ref{sec:results}). The $l_{\rm
    proj} = 0.8$ profiles were generated using the ``thin'' atmosphere
  approximation while the $l_{\rm proj} = 1.09$ profiles were
  generated using the ``extended'' atmosphere approximation (see
  text); in the bottom panels the $l_{\rm proj} = 1.08$ profiles are
  shown for comparison as black/dotted lines.}
\label{fig:initial}
\end{center}
\end{figure*}

For the initial guess above, if $F \ge F_{\rm crit}$ for certain
regions in the atmosphere (see Section~\ref{sub:hydroeq}), we switch
to an ``extended'' atmosphere approximation for those regions. To
obtain an initial guess for extended regions, we assume that the flux
remains at the critical level
\be
F = F_{\rm crit}
\label{eq:Fapprox2}
\ee
throughout the region \citep{paczynski86}. In steady state and
radiative equilibrium (where the material emits as much radiation as
it absorbs), we have from Equation~(\ref{eq:F0thGRTE}) that
\be
\frac{\partial \left(r^2F{\cal R}^2\right)}{\partial r} = 0
\label{eq:dFdr}
\ee
or that
\be
F = \frac{L_{\rm proj}}{4\pi r^2}\left(\frac{{\cal R}_{\rm base}}{{\cal R}}\right)^2
\label{eq:F}
\ee
[cf.\ Equation~(\ref{eq:Lproj})]. As in the thin-atmosphere case, in
steady state with $P_{rr} = u/3$, isotropic $\kappa_\nu^{\rm tot}$ and
$j_\nu^{\rm tot}$, and $u + P_{rr} \ll \rho c^2$, we have from
Equations~(\ref{eq:FA1stGRTE}) and (\ref{eq:Fapprox2}) that
\be
\frac{1}{3}\frac{\partial u}{{\cal V}\partial r} \equiv = \frac{4a}{3}\frac{\partial T_r}{{\cal V}\partial r} = -\rho g_r \,.
\label{eq:dudr2}
\ee
We use the above assumptions, Equation~(\ref{eq:kpapprox}) as an
approximation for $\kappa_F^{\rm tot}$, and $T = T_r$
[cf.\ Equation~(\ref{eq:TTr})], along with
Equations~(\ref{eq:Fapprox2}) and (\ref{eq:F}), to solve for $T_r$ and
$T$ as functions of $r$; and then Equation~(\ref{eq:dudr2}) to solve
for $\rho$ as a function of $r$. Note that
Equations~(\ref{eq:Fapprox2}) and (\ref{eq:F}) combined give
\be
\kappa_F^{\rm tot} = \frac{4\pi cGM}{L_{\rm proj}} {\cal V}\left(\frac{{\cal R}}{{\cal R}_{\rm base}}\right)^2 \propto \frac{1}{1+z} \,;
\label{eq:kpcrit}
\ee
that is, $\kappa_F^{\rm tot}$ must decrease toward the base of the
atmosphere, since the redshift factor $1+z$ is largest there. This
scaling is possible because the total scattering cross section drops
toward the inner, hotter part of the atmosphere according to
Equation~(\ref{eq:kpapprox}). Solving Equation~(\ref{eq:kpcrit}) for
$T$ yields temperature profiles similar in shape to those in figure~1
of \citet{paczynski86}. In the very outer layers of the atmosphere, if
$T$ found in this manner is less than $T$ given by
Equation~(\ref{eq:TTr}), we revert to the ``thin'' atmosphere method
outlined earlier [but with $F$ given by Equation~(\ref{eq:F})
  now]. Example initial guesses are given in Figure~\ref{fig:initial}
for both a thin atmosphere and an atmosphere with an extended
region. As can be seen in the figure, our initial guesses tend to
overestimate the end-of-calculation temperatures and densities. This
is particularly true for extended atmospheres, where the flux at any
point in the atmosphere is close to the critical value such that small
changes in $l_{\rm proj}$ lead to large changes in the temperature and
density profiles. For example, in the solar, $g_{\rm base} =
10^{14}~{\rm cm~s^{-2}}$, $l_{\rm proj} = 1.09$ model our $l_{\rm
  proj} = 1.08$ initial guess is a much better fit to the results of
Section~\ref{sec:results} (differing in $T_r$ by $<10\%$ in most
places) than our $l_{\rm proj} = 1.09$ initial guess (differing in
$T_r$ by $>20\%$ in most places).

\subsection{Monte Carlo method}
\label{sub:MC}

The radiation transfer and material energy equations are solved using
an implicit Monte Carlo (IMC) method; we discuss the basics of the
method here but refer to the original paper
(\citealt{fleckcummings1971}; see also \citealt{wollaber08}) for
details. We also discuss here how we implement general relativistic
effects within the method. In the Appendix we discuss our Compton
scattering algorithm and our treatment of stimulated scattering.

The material quantities are defined on a one-dimensional grid of cells
and change at the end of each time step, as discussed above; but the
radiation field is represented by discrete, stochastic particles that
evolve continuously in space and time and can travel in any
direction. Each Monte Carlo particle represents a packet of photons
that are emitted at approximately the same radius and time, and are
all moving in approximately the same direction with approximately the
same frequency. Note that we do not localize these packets in polar
angle or azimuthal angle; instead, we exploit the spherical symmetry
of the model by grouping all similar photons across the entire $4\pi$
steradians of (spatial) solid angle into one ``packet''. Therefore, we
do not track location angles for the particles (only direction angles;
see below).

Because the material quantities are only updated at the end of the
time step but the radiation field evolves continuously in time,
beginning-of-time-step values for, e.g., the material temperature are
used during radiation-matter interactions (photon emission,
absorption, and scattering). Solving Equation~(\ref{eq:GRTE}) as it is
written, i.e., explicitly, can lead to large fluctuations in the
radiation and material energies. The solution proposed by
\citet{fleckcummings1971} is to treat the radiation transfer equation
(Section~\ref{sub:rteq}) in a more implicit manner, by using the
end-of-time-step material temperature $T^{n+1}$ for $B_\nu(T)$ in the
equation, rewriting the equation in terms of the
beginning-of-time-step temperature value, and then solving the
resulting equation using Monte Carlo techniques. The expression for
$T^{n+1}$ is found from the material energy equation
(Section~\ref{sub:meeq}) assuming $I_\nu$ is given by its current
value (not its average value over the time step) and that scattering
is isotropic. Following their method, we obtain a new version of
Equation~(\ref{eq:GRTE}) that differs only in the $j_\nu^{\rm tot}$
term [cf.\ Equation~(\ref{eq:jnu})], now given by
\be
j_\nu^{\rm tot} = f_{\rm abs}j_\nu + (1-f_{\rm abs})j_\nu^{\rm eff} + j_\nu^{\rm sc}(\hat{\Omega}) \,,
\label{eq:jnu3}
\ee
where
\be
j_\nu^{\rm eff} = \chi_\nu \rho \int_0^\infty d\nu' \kappa_{\nu'} \int_{4\pi} d\Omega' I_{\nu'}(\hat{\Omega})
\ee
is the effective scattering coefficient,
\be
\chi_\nu = \frac{\kappa_\nu B_\nu(T)}{\int_0^\infty d\nu' \kappa_{\nu'} B_{\nu'}(T)}
\ee
is the reemission spectrum,
\be
f_{\rm abs} = \frac{1}{1+(4aT^3/\rho c_V) c \Delta t_{\rm cell} \rho\kappa_{\rm P}}
\ee
is the effective absorption factor and
\be
\kappa_{\rm P} = \frac{\int_0^\infty d\nu \kappa_\nu B_\nu(T)}{\int_0^\infty d\nu B_\nu(T)}
\ee
is the Planck-averaged absorption opacity. In the IMC method, a
fraction $f_{\rm abs}$ of the total emission in a time step is handled
in the standard way, by creating new particles and decreasing the cell
energy (see below); but the remaining $1-f_{\rm abs}$ is handled
through ``effective scattering'': a particle undergoing effective
scattering is given a new direction, uniformly random, and a new
frequency, random but weighted by $\chi_\nu$. Physically, the material
absorbs the particle and then shortly after reemits the absorbed
energy isotropically. Numerically, the time step is too large for the
absorption and emission processes to be resolved separately, such that
they appear as one scattering process ($f_{\rm abs} \rightarrow 0$ as
$\Delta t_{\rm cell} \rightarrow \infty$).

The simulation is initially populated with Monte Carlo particles
according to Equation~(\ref{eq:unu}). During each time step, the code
creates new Monte Carlo particles to simulate the process of
emission. The number of particles created depends on how many
particles are already in the simulation: if there are relatively few
existing particles then many are created, but if the number of
existing particles is close to the maximum allowed value then few are
created. We typically set the maximum to $10^6$ particles; we want
enough particles to populate every cell and frequency group
(Section~\ref{sub:model}), but not so many that our simulation runs
slowly. For each particle, the details of its creation (location,
direction, etc.)\ are randomly chosen. The cell of creation is randomly
chosen with weighting $\Delta U_{\rm cell}$, the amount of energy the
cell emits in the time step. Using Equations~(\ref{eq:kirchoff}),
(\ref{eq:F0thGRTE}), and (\ref{eq:jnu3}) we have
\be
\Delta U_{\rm cell} = f_{\rm abs} c \Delta t_{\rm cell} V_{\rm cell}\rho \kappa_{\rm P} aT^4 \,,
\label{eq:Du}
\ee
where $V_{\rm cell}$ is the volume of the cell and $\Delta t_{\rm
  cell}$ is the time step for the cell (see below). For simplicity,
the radius within the cell and the time within the time step are
randomly chosen with uniform weighting; from
Equation~(\ref{eq:kirchoff}), the direction is weighted uniformly
(isotropically) while the frequency is weighted by $\kappa_\nu
B_\nu(T)$. The particle also has an energy weight, equal to the total
energy of the photon packet. This is not randomly chosen; instead, all
of the particles ``emitted'' in a time step have the same energy
weight, given by
\be
w_{\rm emit} = \frac{\sum_{\rm cells} \Delta U_{\rm cell}}{N_{\rm emit}}
\ee
with $N_{\rm emit}$ the number of Monte Carlo particles emitted in
that step. At the end of the time step the cell temperature is
adjusted due to emission using
\be
\Delta T_{\rm emit} = -\frac{\Delta U_{\rm cell}}{V_{\rm cell}\rho c_V}
\label{eq:Temit}
\ee
(see below).

Particles are also created at the inner boundary of the simulation,
the base of the atmosphere. Here the physical process for the creation
of the particles is not emission, but the escape of photons from the
hot layers beneath the atmosphere. As before, the details of creation
are chosen randomly; the time within the time step is chosen with
uniform weighting, while from Equation~(\ref{eq:flux}), the direction
is weighted by $\mu$ and the frequency is weighted by $B_\nu$. The
energy weight for these particles is
\be
w_{\rm base} = \frac{\Delta U_{\rm base}}{N_{\rm base}}
\ee
with $N_{\rm base}$ the number of particles created at the base in the
time step and $\Delta U_{\rm base}$ the energy entering the simulation
from the base of the atmosphere. From Equations~(\ref{eq:flux}) and
(\ref{eq:Ibase}) we have
\be
\Delta U_{\rm base} = \Delta t_{\rm base} A_{\rm base} \sigma_{\rm SB} T_{\rm base}^4 \,,
\label{eq:Dubase}
\ee
where $A_{\rm base}$ is the area of the base and $\Delta t_{\rm cell}$
is the time step at the base [cf.\ Equation~(\ref{eq:u})].

The created particles (photon packets) are transported according to
the left-hand side of Equation~(\ref{eq:GRTE}). If general
relativistic effects are ignored, the particles travel in straight
lines. For a particle traveling a distance $s$ from the position
$(r_{\rm old},\theta_{\rm old},\phi_{\rm old})$ and moving in the
direction $\hat{\Omega}$ initially defined by $\hat{\Omega} =
\Omega_{r,\rm old}\hat{r}_{\rm old} + \Omega_{\theta,\rm
  old}\hat{\theta}_{\rm old}$, where $\Omega_{r,\rm old} \equiv
\mu_{\rm old}$, the new radius of the particle is given by the law of
cosines:
\be
r_{\rm new} = \sqrt{s^2 + r_{\rm old}^2 + 2sr_{\rm old}\mu_{\rm old}} \,.
\label{eq:rnew}
\ee
The new polar angle for the direction could be found in a similar
manner, using Equation~(\ref{eq:rnew}) and the law of cosines for
$r_{\rm old}$ in terms of $s$ and $r_{\rm new}$: $\mu_{\rm new} = (s +
r_{\rm old}\mu_{\rm old})/r_{\rm new}$. However, in our code this
quantity is solved in a manner that is more consistent with general
relativity, as we describe below. Note that because of the spherical
symmetry of the problem, we orient the direction $\hat{\Omega}$ such
that it has no azimuthal component $\Omega_\phi$ and do not track the
azimuthal angle for the position (or the polar angle for the position,
once $\mu_{\rm new}$ is known).

After the particle is transported using Equation~(\ref{eq:rnew}), we
apply general relativistic effects. Due to the gravitational redshift
term in Equation~(\ref{eq:GRTE}) [the one proportional to
  $(g_r/c^2)(\partial (I_\nu\nu^{-3})/\partial \nu)$], the new
frequency and weight of the particle are given by
\be
\nu_{\rm new} = \nu_{\rm old} \frac{{\cal R}_{\rm old}}{{\cal R}_{\rm new}}
\label{eq:nunew}
\ee
and
\be
w_{\rm new} = w_{\rm old} \frac{{\cal R}_{\rm old}}{{\cal R}_{\rm new}} \,;
\ee
the number of photons in the packet remains the same but their energy
changes. Time dilation [represented by the $(1/c)(\partial u/{\cal
    R}\partial t)$ term in Equation~(\ref{eq:GRTE})] is taken into
account by using a different
\be
\Delta t_{\rm cell} = \Delta t_{\rm base} \frac{{\cal R}}{{\cal R}_{\rm base}}
\ee
for each cell. For simplicity we consider gravitational light bending
[represented by the $(g_r/c^2)(\partial I_\nu/\partial \mu)$ term] as
a perturbation on the particle trajectory; see
Section~\ref{sub:gravity}. The trajectory of a photon traveling in a
Schwarzschild geometry is given by \citep[e.g.,][]{misner73:_gravit}
\be
\frac{d\theta}{dr} = -\frac{b}{r\sqrt{r^2-b^2\left(1-2GM/c^2r\right)}} \,,
\label{eq:GRtraj}
\ee
where $b$ is the impact parameter. The impact parameter can be found
by using the pre-transport position and direction quantities ($r_{\rm
  old}$, $\Omega_{r,\rm old}$, and $\Omega_{\theta,\rm old}$) to
equate Equation~(\ref{eq:GRtraj}) and
\be
\frac{d\theta}{dr} = -\frac{\Omega_\theta}{r\Omega_r} \,,
\label{eq:domegadr}
\ee
which describes the local propagation direction of the particle. If
$\hat{\Omega} = \pm \hat{\theta}$ [such that $\Omega_r = 0$ and
  Equation~(\ref{eq:domegadr}) is undefined], we set $b = \pm r_{\rm
  old}/\sqrt{1-2GM/c^2r_{\rm old}}$. We then solve for $\mu_{\rm new}
\equiv \Omega_{r,\rm new}$ by equating Equations~(\ref{eq:GRtraj}) and
(\ref{eq:domegadr}) with $r_{\rm new}$ from
Equation~(\ref{eq:rnew}). Our approximation of using the Newtonian
$r_{\rm new}$ rather than a post-transport radius calculated
self-consistently with Equation~(\ref{eq:GRtraj}) is accurate to first
order in $(2GM/c^2r) (\Delta r/r)$ (Section~\ref{sub:gravity}).

After the particle is transported and general relativistic effects are
applied, it undergoes an event. There are five particle events that
our code accounts for: 1) the particle is absorbed; 2) the particle is
effectively scattered (see earlier in this section); 3) the particle
is actually scattered; 4) the particle reaches the boundary of a cell;
5) the time step ends. Each of these events has a distance associated
with it, e.g., for event 4, the distance the particle needs to travel
to reach the boundary of the cell. For each particle, the code
determines which of the events has the shortest distance (i.e., which
event will happen first), transports the particle by that distance
[$s$ in Equation~(\ref{eq:rnew})], and then carries out the event as
described below.

For event 1: absorption and event 2: effective scattering, the
distance traveled is given by
\be
s_{\rm abs} = -\frac{\ln\xi}{\rho\kappa_\nu} \,,
\label{eq:dabs}
\ee
where $\xi$ is a (uniformly distributed) random number between 0 and
1. Equation~(\ref{eq:dabs}) comes from assuming a probability for
photon absorption of $e^{-\rho\kappa_\nu s}$
\citep{fleckcummings1971}. A fraction $f_{\rm abs}$ of the particles
undergo absorption, in which case the location of the event and the
energy weight of the particle are recorded and then the particle is
destroyed; the remainder undergo effective scattering, in which case
the direction and frequency are adjusted as described earlier in this
section and the energy weight remains at $w_{\rm new}$. For both
events, the change in the radial momentum of the particle is recorded
using only the pre-scattering contribution:
\be
\Delta p_{r,\rm abs} = - \frac{\mu_{\rm new} w_{\rm new}}{c} \,;
\label{eq:prabs}
\ee
we do not include the post-scattering contribution $\mu'_{\rm new}
w_{\rm new}/c$ in $\Delta p_{\rm abs}$ here, since due to the
isotropic nature of effective scattering this contribution is on
average zero (we do not record the momentum of emitted particles
either, for the same reason). For event 3: actual scattering, the
distance traveled is given by Equation~(\ref{eq:dabs}) but with
$\kappa_\nu^{\rm sc}$ in place of $\kappa_\nu$. The direction and
frequency of the particle are adjusted based on the (Compton)
scattering opacity in Equation~(\ref{eq:kappascat2}), as described in
the Appendix \citep[cf.][]{canfieldetal1987}, the energy weight is
adjusted according to
\be
w'_{\rm new} = \frac{\nu'_{\rm new}}{\nu_{\rm new}}w_{\rm new} \,,
\ee
and then the location and change in energy weight is recorded. The
change in momentum is recorded using
\be
\Delta p_{r,\rm sc} = \frac{\mu'_{\rm new} w'_{\rm new} - \mu_{\rm new} w_{\rm new}}{c} \,.
\label{eq:prsc}
\ee
For event 4: cell boundary, the distance traveled depends on the
particle direction. The first cell boundary reached will be the inside
boundary $r_{\rm in}$ if (using the Pythagorean theorem)
\be
r_{\rm old}\mu_{\rm old} \le -\sqrt{r_{\rm old}^2-r_{\rm in}^2} \,,
\ee
otherwise it will be the the outside boundary $r_{\rm out}$. Then the
distance traveled is given by the law of cosines (ignoring light
bending; see above):
\be
s_{\rm bnd} =
\left\{
\begin{array}{ll}
-r_{\rm old}\mu_{\rm old}+\sqrt{r_{\rm old}^2\mu_{\rm old}^2+r_{\rm out}^2-r_{\rm old}^2} \,, & \rm{outside}; \\
-r_{\rm old}\mu_{\rm old}-\sqrt{r_{\rm old}^2\mu_{\rm old}^2+r_{\rm in}^2-r_{\rm old}^2} \,, & \rm{inside}.
\end{array}
\right.
\label{eq:dbnd}
\ee
If the particle leaves the simulation through the inner boundary it is
destroyed. If the particle leaves the simulation through the outer
boundary, its energy weight is recorded and then the particle is
destroyed. For event 5: end of time step, the distance traveled is
given by the speed of light multiplied by the time remaining until the
end of the time step. After the particle is propagated to event 5, its
properties (location, direction, frequency, and energy weight) are
stored in memory until the next time step.

After the event, if the particle has not escaped the simulation or
been absorbed and there is still time remaining in the current time
step, the process continues with another event. At the end of the time
step, the energy in each cell is decreased by the combined energy
weights of all particles created in the cell during the step (through
emission; see above), increased by the energy weights of all particles
destroyed in the cell during the step (through absorption), and
modified by all energy weight changes (due to scattering);
cf.\ Equation~(\ref{eq:Temit}). The fluence (the time- and
area-integrated flux) at the outer boundary of the simulation ${\cal
  F}_{\rm surf}$ is increased by the quantity $w/c$ for all particles
that cross the boundary, while the group-dependent fluence ${\cal
  F}_{\rm g,surf}$ is increased by $w/c$ only for those particles with
frequencies in the range of the group. Here
\be
{\cal F}_{\rm surf} = \sum_{\rm groups} {\cal F}_{\rm g,surf}
\ee
and
\be
{\cal F}_{\nu,\rm surf} \simeq {\cal F}_{\rm g,surf}/\Delta\nu_{\rm g} \,,
\ee
where ${\rm g}$ is the frequency group and $\Delta\nu_{\rm g}$ is the
width of that frequency group. The outgoing spectral flux at the
surface $F_{\nu,\rm surf}$ and the luminosity are given by
\be
F_{\nu,\rm surf} = \frac{\Delta {\cal F}_{\rm g,surf}/\Delta\nu_{\rm g}}{A_{\rm surf}\Delta t_{\cal F}}
\label{eq:Fnug}
\ee
and
\be
L_{\rm surf} = \frac{\Delta {\cal F}_{\rm surf}}{\Delta t_{\cal F}} \,.
\ee
Here $\Delta t_{\cal F}$ is the time interval over which $\Delta {\cal
  F}_{\rm surf}$ is taken; to reduce noise, we typically use $\Delta
t_{\cal F} = 10^{-6}~{\rm s} \simeq 1000 \Delta t_{\rm surf}$. Note
that in the Monte Carlo method, the flux measured by counting
particles crossing a cell boundary is naturally the flux in the
direction normal to the boundary; there is no need to multiply each
particle's energy weight by $\mu$
[cf.\ Equation~(\ref{eq:flux})]. This is because each particle
implicitly carries with it a solid angle $\Delta \Omega$, such that
when the particle intersects the boundary the energy weight is spread
out over an area proportional to $\mu$.

\subsection{Method for ensuring hydrostatic balance}
\label{sub:hydrosolve}

\begin{figure}
\begin{center}
\includegraphics[width=\columnwidth]{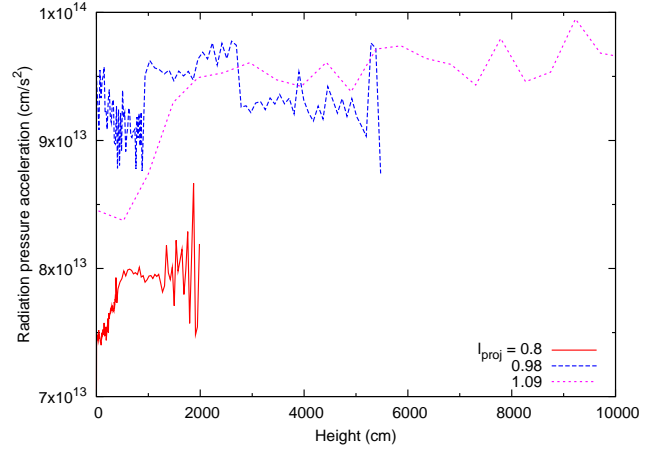}
\caption{The quantity $\Delta p_{r,\rm cell}/\rho V_{\rm cell}\Delta
  t_p$ (where $\Delta p_{r,\rm cell}$ is the momentum deposited into
  the material; see text) as a function of height above the base of
  the atmosphere, for atmosphere models with solar composition,
  $g_{\rm base} = 10^{14}~{\rm cm~s^{-2}}$, and a variety of
  luminosity ratios.}
\label{fig:gRad}
\end{center}
\end{figure}

Every several time steps we adjust the structure of the atmosphere to
maintain hydrostatic balance. We keep $\rho_{\rm base}$ fixed and
adjust the mass density in the cells, attempting to satisfy
Equation~(\ref{eq:hydroeq}) in every cell. For stability reasons, we
typically choose an interval between adjustments of
$5\times10^{-6}~{\rm s} \simeq 5000$ time steps, such that the
atmosphere is close to radiative equilibrium, and restrict the density
change in any particular cell to ten percent or less in each
adjustment. The radiation term in Equation~(\ref{eq:hydroeq}) is given
by
\bal
{}& -\frac{1}{c} \int_0^\infty d\nu \int_{4\pi} d\Omega \mu \left[-\rho\kappa_\nu^{\rm tot}(\hat{\Omega}) I_\nu(\hat{\Omega}) + j_\nu^{\rm tot}(\hat{\Omega})\right] \nonumber\\
{}& \qquad\qquad\qquad\qquad = \frac{\Delta p_{r,\rm cell}}{V_{\rm cell}\Delta t_p} \,,
\label{eq:radbalance}
\eal
where $\Delta t_p$ is the time interval over which $\Delta p_{r,\rm
  cell}$ is taken. To reduce noise, we typically use $\Delta t_p =
5000 \Delta t_{\rm cell}$; i.e., the momentum is differenced over the
entire interval between density adjustments. Note that because of the
symmetry of the model, in Equation~(\ref{eq:radbalance}) we ignore
contributions to the radiation term due to momentum changes in
non-radial directions. Figure~\ref{fig:gRad} shows the ``radiation
pressure acceleration'' \citep[cf.\ equation~7 of][]{spw11}, given by
the quantity $\Delta p_{r,\rm cell}/\rho V_{\rm cell}\Delta t_p$ from
Equation~(\ref{eq:radbalance}), which should be less than or equal to
$g_r$ when the atmosphere is in hydrostatic balance.

\subsection{Reaching the end-of-calculation equilibrium}
\label{sub:equilibrium}

\begin{figure}
\begin{center}
\includegraphics[width=\columnwidth]{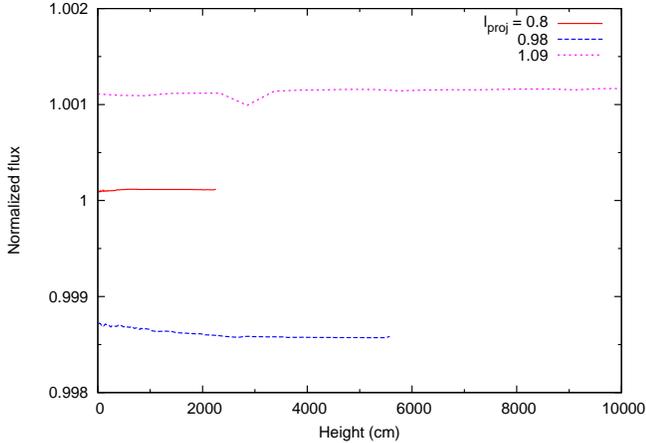}
\caption{The quantity $4\pi r^2F{\cal R}^2/(l_{\rm proj}L_{\rm
    Th}{\cal R}_{\rm base}^2)$ (see text) as a function of height
  above the base of the atmosphere, for the same atmosphere models as
  in Figure~\ref{fig:gRad}.}
\label{fig:radeq}
\end{center}
\end{figure}

We continue the process described in Sections~\ref{sub:MC} and
\ref{sub:hydrosolve}, i.e., every time step transporting the photons
and coupling them to the material in a manner described by
Equations~(\ref{eq:GRTE}) and (\ref{eq:MEE}), and then every several
time steps partially restoring hydrostatic balance using
Equation~(\ref{eq:hydroeq}), until steady state and radiative
equilibrium is reached. Our criterion for reaching steady state is
that the luminosity $L_{\rm proj}$ changes by less than one part in
$10^6$ per time step. This criterion is typically satisfied after
around $10^6$ time steps for thin atmospheres, more for thick
atmospheres or when the system is far from hydrostatic balance. We do
not explicitly check whether radiative equilibrium is reached;
however, once the luminosity meets the above criterion, the quantity
$r^2F{\cal R}^2$ usually varies by no more than 1\% across the entire
atmosphere [cf.\ Equation~(\ref{eq:dFdr})]. We then adjust $T_{\rm
  base}$ as described in Section~\ref{sub:model}, and repeat the
process to reach a new steady state/radiative equilibrium. We let the
simulation run until the projected luminosity is within 0.2\% of the
desired value $l_{\rm proj} L_{\rm Th}$. At this point the simulation
is very nearly in radiative equilibrium, as is shown in
Figure~\ref{fig:radeq}.
 
\section{Effect of various physical processes on the observed spectra}
\label{sec:physics}

The pieces of physics we include in our simulations have varying
degrees of effect on our results. Here we describe each piece.

\subsection{Radiation transport}
\label{sub:radtrans}

In the outer layers of a hot neutron star ($\rho \alt 1~{\rm
  g~cm^{-3}}$, $T \agt 10^7$~K), radiation transport is the dominant
form of energy transport; heat conduction and convection contribute
very little \citep{joss77,rajagopal96,potekhin97,weinbergetal2006}. We
therefore only consider radiation transport in our calculations
here. Since we are modeling the deviation of the outgoing radiation
spectra from a Planck function (i.e., from blackbody radiation), our
problem is inherently multi-frequency. We assume a spherically
symmetric atmosphere and a spherically symmetric but anisotropic
radiation field (see Section~\ref{sec:method}). Axially symmetric or
fully three-dimensional atmospheres are more accurate if the neutron
star is highly non-spherical (due to rotation) or the thermonuclear
burning powering the burst is not uniform over the star (see
\citealt{miller2013} and references therein), but they are
computationally very expensive.

An anisotropic radiation field is necessary to correctly model the
optically thin, outer atmosphere
\citep[e.g.,][]{rybickilightman1986}. For example, we find that the
radiation energy density in the outer layers is typically a few
percent lower when using anisotropic radiation transport than when
using radiation diffusion. Note that while the Monte Carlo method has
many advantages, including the automatic treatment of multi-frequency
and anisotropic physics mentioned above, one key disadvantage is
stochastic noise. The effect of this noise can be seen in many of the
figures in this paper; it is most noticeable at the high-$r$ end of
the material temperature profiles (e.g., Figure~\ref{fig:initial}) and
at the high- and low-frequency ends of the spectra (e.g.,
Figure~\ref{fig:thomsonCompton}), where there are fewer particles. We
have run our simulations with enough particles to compensate for this
noise, but the computation time is an order of magnitude longer than
for a diffusion calculation with a converged spectrum.

\subsection{Radiation-material interaction}
\label{sub:radmat}

For a review on interactions between radiation and material, see
\citet{castor2004}. We assume that the radiation at the base of the
atmosphere is in thermal equilibrium with the material there. For this
to be an accurate assumption, most photons emitted at the base of the
atmosphere must be absorbed before they escape; this latter condition
occurs when \citep{rybickilightman1986}
\be
\sqrt{\tau_{\nu,\rm base} \tau_{\nu,\rm base}^{\rm tot}} \agt 1 \,,
\label{eq:taueff}
\ee
where $\tau_{\nu,\rm base}$ is the absorption-only optical depth at
the base of the atmosphere and $\tau_{\nu,\rm base}^{\rm tot}$ is the
total (absorption plus scattering) optical depth
[cf.\ Equation~(\ref{eq:dtaudr})]. For the highest frequencies of
interest here \citep[$h\nu \sim 30$~keV; cf.][]{spw11}, the absorption
opacities are very small, such that satisfying
Equation~(\ref{eq:taueff}) requires frequency-averaged total optical
depths $\tau_{\rm base}^{\rm tot} \agt 100$. Ideally we should place
the atmosphere base deep enough that $\tau_{\rm base}^{\rm tot} \gg
100$, for greater accuracy; for Monte Carlo calculations, however, the
computation time increases quadratically with total optical depth
\citep{densmore07} and so we place the base right at this critical
value. We have performed convergence studies in $\tau_{\rm base}^{\rm
  tot}$ for a few cases and found differences of only a couple percent
between the spectra in the $\tau_{\rm base}^{\rm tot} = 100$ case and
the converged answer, and no detectable difference between the color
correction factors for the two cases.

A change to the radiation field in the optically thick, inner layers
of the atmosphere diffuses outward at a velocity $v_{\rm diff} \simeq
c/\tau^{\rm tot}$, such that it reaches the surface after a time
$t_{\rm diff} \simeq H\tau^{\rm tot}/c$, where $H$ is the thickness of
the atmosphere; for $\tau_{\rm base}^{\rm tot} = 100$ we have that
thin atmospheres reach radiative equilibrium in $\sim 10^{-5}$~s,
while extended ($\sim 10$~km) atmospheres reach radiative equilibrium
in $\sim 10^{-2}$~s. For comparison, the temperature at the base of
the atmosphere, which tracks the X-ray burst evolution, changes on a
time scale of several seconds (the burst time). We therefore treat the
temperature at the base of the atmosphere as constant over our
simulation time and evolve the system to steady state
\citep[cf.][]{shaposhnikov02}.

We consider both scattering and absorption-emission of photons by
electrons. Specifically, we include Compton scattering -- scattering
of photons by free electrons (or ``nearly free'' electrons; see
Section~\ref{sub:opacity}); inverse bremsstrahlung -- absorption of
photons by free electrons in the presence of an ion -- and
photoionization/photoexcitation -- bound-free/bound-bound transitions
of electrons through absorption of photons; and the reverse processes
(that is, emission of photons by free or bound electrons). The exact
implementation of each of these radiation-material interactions is
discussed in Section~\ref{sub:opacity}.

\begin{figure}
\begin{center}
\includegraphics[width=\columnwidth]{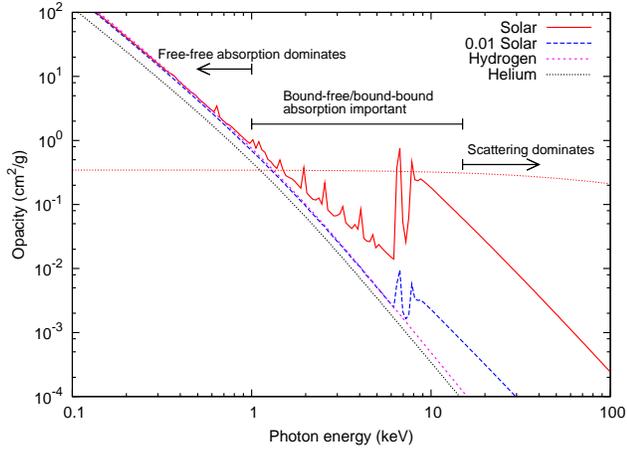}
\caption{Absorption opacity as a function of photon energy, for $\rho
  = 0.46$~\cmsqperg and $k_{\rm B}T = 1$~keV. The opacities shown here
  are averaged over frequency group; in our simulation we use 300
  logarithmically spaced groups in the the range $\nu = 10$~eV to
  $1$~MeV (see text). Each model curve is labeled with the atmosphere
  composition for that model. For the ``Solar'' composition model, the
  scattering opacity ignoring stimulated scattering
  [cf.\ Equation~(\ref{eq:kappascat})] is also shown for comparison,
  as a thin solid line. The arrows show the approximate frequency
  regions where each type of absorption or scattering is important for
  these models.}
\label{fig:opacity}
\end{center}
\end{figure}

\begin{figure}
\begin{center}
\includegraphics[width=\columnwidth]{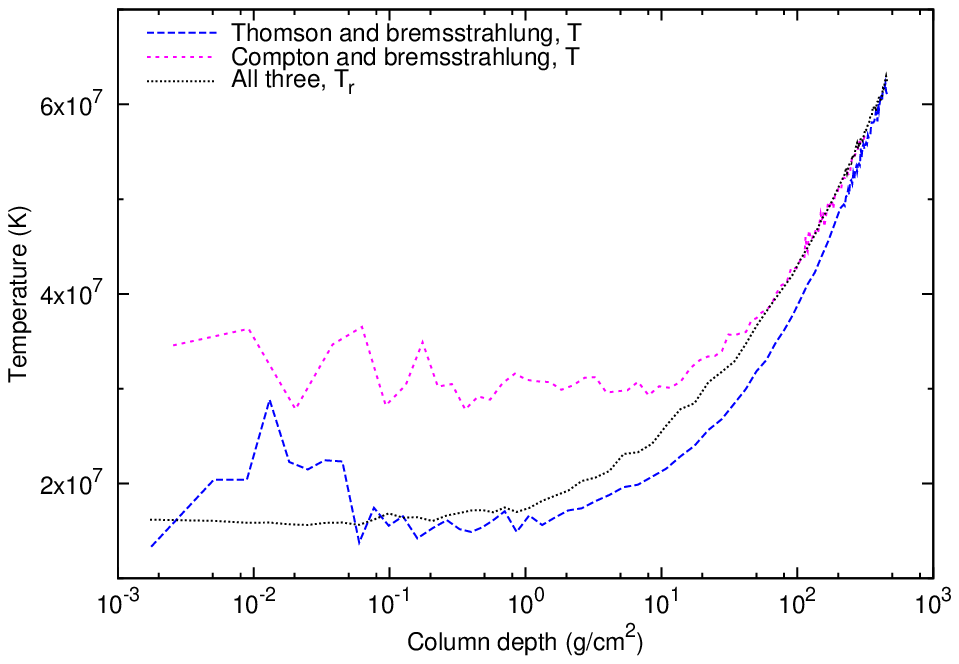}
\includegraphics[width=\columnwidth]{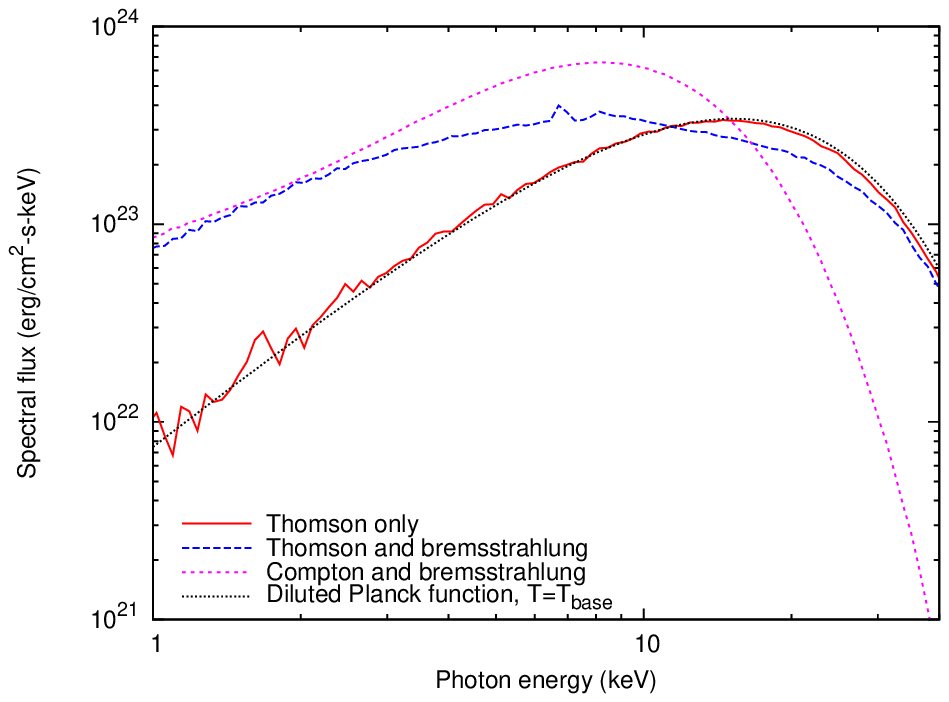}
\caption{Material and radiation temperature profiles (top panel) and
  the outgoing radiation spectrum (bottom panel), for a
  solar-composition, $g_{\rm base} = 10^{14}~{\rm cm~s^{-2}}$, $l_{\rm
    proj} = 0.98$ atmosphere. The three line colors/patterns represent
  three different models for scattering and absorption: red/solid
  curves represent Thomson scattering only, blue/long-dashed curves
  represent Thomson scattering and bremsstrahlung, and
  purple/short-dashed curves represent Compton scattering and
  bremsstrahlung. In the top panel, we do not show a curve of the
  material temperature $T$ for the model with Thomson scattering only,
  since the material has no effect on the outgoing spectrum in this
  case. In addition, we only show one representative curve of the
  radiation temperature $T_r$ (as a black/dotted line), since the
  $T_r$ curves are very similar for all three models. In the bottom
  panel, we show for comparison the diluted Planck function $\epsilon
  B_\nu(T_{\rm base})$ with $\epsilon = F/\sigma_{\rm SB} T_{\rm
    base}^4$ (see text).}
\label{fig:thomsonCompton}
\end{center}
\end{figure}

Figure~\ref{fig:opacity} shows the contributions of the scattering and
absorption processes to the total frequency-dependent opacity. Inverse
bremsstrahlung is the main contributor to the total opacity at low
frequencies, because of the $\nu^{-3}$ dependence of this
process. Compton scattering dominates at higher frequencies, except at
low densities and temperatures for solar-like compositions, where
photoionization and photoexcitation dominate the $h\nu \sim
1$--$10$~keV range. In terms of the outgoing radiation spectra,
photoionization and photoexcitation cause absorption lines and edges
for metal-rich bursts at low luminosities ($l_{\rm proj} \alt 0.01$,
see Figure~\ref{fig:spwSpecComp}), but have little effect at high
luminosities. Bremsstrahlung/inverse bremsstrahlung and Compton
scattering, on the other hand, have strong effects at all compositions
and luminosities we consider in this paper. Compton scattering is also
the main contributor to the optical depth in the atmosphere, and
therefore sets the general structure of the radiation there. To
highlight the effects of these latter two processes we compare here
three atmosphere models, shown in Figure~\ref{fig:thomsonCompton}: one
with Thomson scattering (the non-relativistic limit of Compton
scattering) only, one with Thomson scattering and bremsstrahlung, and
one with Compton scattering and bremsstrahlung
\citep[cf.][]{madej04,miller11}. The top panel of
Figure~\ref{fig:thomsonCompton} shows the material and radiation
temperatures $T$ and $T_r$ as a function of column depth for each
model, while the bottom panel shows the outgoing
spectrum.\footnote{Note that the $\tau_{\rm base}^{\rm tot} \agt 100$
  convergence discussed earlier in this section only holds for models
  that include Compton scattering. The simplified, Thomson-scattering
  models discussed here are not fully converged until $\tau_{\rm
    base}^{\rm tot} \gg 100$. Since we did not extend our models to
  such large depths, the results presented in
  Figure~\ref{fig:thomsonCompton} should be used for qualitative
  comparisons only.}

In the Thomson-scattering-only atmosphere, the radiation field near
the surface differs in magnitude from that at the base, but has the
same frequency distribution; i.e., the radiation field is represented
by a Planck function at the base [$I_\nu(\hat{\Omega}) = B_\nu(T_{\rm
    base})$] and a diluted Planck function at the surface
[$I_\nu(\hat{\Omega}) = \epsilon(\hat{\Omega})B_\nu(T_{\rm base})$,
  where $\epsilon(\hat{\Omega})$ is the frequency-independent dilution
  factor]. This is because, although Thomson scattering reduces the
number of photons traveling outward from the base, the direction and
amount of scatter are independent of frequency and no energy is
transferred between the photons and the material. For this type of
atmosphere the material and the radiation are decoupled; i.e., $T$ and
$T_r$ are independent of each other.

In the bremsstrahlung-plus-Thomson atmosphere, due to the strong
frequency dependence of the bremsstrahlung process, low-frequency
photons are absorbed after traveling only a short distance, and those
that reach the surface are emitted from nearby, relatively cool
layers. Hence, in this atmosphere the radiation field at low
frequencies is described by (when the material is in LTE) $I_\nu =
B_\nu(T)$, where $T$ is the local material temperature. Mid-frequency
photons are emitted from deeper, hotter layers, and are described by
$I_\nu = B_\nu(T_\nu^*) > B_\nu(T)$ with $T_\nu^* > T$ that increases
with frequency. The highest-frequency photons are neither absorbed nor
emitted, but are scattered up from the base; the radiation field at
these frequencies is $B_\nu(T_b)/A \gg B_\nu(T)$, where $T_b$ is the
temperature at the base of the atmosphere and $A$ is a normalization
constant. The total radiation energy density is therefore larger than
$(1/c)\int d\nu \int d\Omega B_\nu(T) = aT^4$, such that $T_r > T$
\citep[cf.][]{mihalas78}. The effect is strongest in the outer layers,
where the radiation is farthest from thermal equilibrium. At the
surface and in steady state, $T \sim 0.5 T_r$ for this type of
atmosphere.

Compton scattering modifies this picture by transferring energy
between the radiation field and the material with each scatter. During
a scattering event, the fraction of energy gained by the photon is
\be
\frac{\Delta\nu}{\nu} = \frac{4k_{\rm B}T-h\nu}{m_ec^2} \,,
\ee
assuming the electrons in the material are in thermal equilibrium
\citep{rybickilightman1986}. Low-to-mid-frequency photons gain a small
amount of energy per scatter ($\Delta\nu \sim \pm0.01\nu$), and
scatter at most a few times before they are absorbed, such that $I_\nu
\simeq B_\nu(T)$ or $B_\nu(T_\nu^*)$ as above; here, however, the
deeper layers are not necessarily hotter than the local
layer. High-frequency photons lose a large amount of energy per
scatter ($\Delta\nu \sim 0.1\nu$) and scatter multiple times, such
that the radiation field at high frequencies is of much lower energy
density than in the Thomson-scattering case. The material temperature
$T$ is therefore larger relative to $T_r$ in this type of atmosphere,
and in the outer layers where the absorption opacity is lowest, the
downscatter effect is so strong that $T > T_r$.

Note that in radiative equilibrium, where the flux through the
atmosphere is constant, the radiation temperature $T_r$ decreases with
increasing radius at a prescribed rate; e.g., in the diffusion
approximation, $F \equiv -(c/\rho\kappa^{\rm tot}) d(aT_r^4)/dr = {\rm
  constant}$. The material temperature $T$ has very little effect on
this rate because it only enters the flux equation through the
opacity, and only very weakly in the scattering-dominated atmospheres
considered here. This is why the models shown in
Figure~\ref{fig:thomsonCompton}, which all have the same atmospheric
flux, have nearly the same $T_r$ profiles. Because the $T_r$ profile
is essentially fixed by the flux $F$ (or $l_{\rm proj}$ in our
formalism; see Section~\ref{sub:model}), the $T$ profile adjusts to
the $T_r$ profile, not the other way around. In the
bremsstrahlung-plus-Thomson atmosphere $T$ drops faster than $T_r$
starting at an optical depth of of unity, and continues to separate
from $T_r$ until $T$ is half $T_r$ at the surface. In contrast, in the
bremsstrahlung-plus-Compton atmosphere $T$ drops faster than $T_r$ in
the mid layers of the atmosphere, but then rises in the outer layers
to be equal to or greater than $T_r$ (in addition to
Figure~\ref{fig:thomsonCompton}, see, e.g., figure~3 of either
\citealt{madej04} or \citealt{spw12}). Effectively, scattering
transports many high-frequency photons to the outer layers and these
photons can not be efficiently absorbed by inverse bremsstrahlung (due
to its $\nu^{-3}$ dependence), so the layers cool off. At the same
time, however, many of these high-frequency photons are downscattered,
causing the outer layers to heat up. The balance between these two
attributes of scattering determines the overall temperature structure
in the atmosphere.

\subsection{Opacity}
\label{sub:opacity}

Our absorption opacities are provided by the LANL TOPS
code\footnote{\url{http://aphysics2.lanl.gov/cgi-bin/opacrun/tops.pl}},
which calculates frequency-averaged opacities from the monochromatic
cross sections in the LANL OPLIB database
\citep{magee95,frey13}. Specifically, we request that the TOPS code
average opacities over each frequency group using a Planck (linear
with weight $B_\nu$) average at the local material temperature $T$. We
have chosen this averaging for simplicity, since the online version of
TOPS provides the absorption opacity in terms of a Planck average and
since using $T$ requires fewer variables in the lookup table (only
$\rho$ and $T$ rather than $\rho$, $T$, and $T_r$). We have checked
for a few cases that using $T_r$ or a Rosseland (inverse with weight
$\partial B_\nu/\partial T$) average instead does not noticeably
change our results; but we do not expect it to due to the narrow
widths of the frequency groups in our simulations. Note that we can
not use a mixture of Rosseland and Planck averages as in
\citet{frey13}, because we are using Monte Carlo transport where there
is no distinction between the opacity used in modeling the transport
of photons and that used in the equilibration of the photons and the
material.

The OPLIB database takes into account the ionization level of the
material, though the atmosphere is mostly ionized at the temperatures
considered here. A plasma cutoff is included in the form of an
artificially large opacity for $h\nu \alt 30 \rho^{1/2}$~eV, though
this cutoff is lower than the lowest frequencies of interest in our
simulations except in the deepest parts of the atmosphere for the
hottest stars. The correction factor for stimulated emission,
$1-e^{-h\nu/k_{\rm B}T}$ [Equation~(\ref{eq:kappa})], is automatically
included in the opacities returned from the database through
TOPS. Thermal broadening is also included in the database.

\begin{figure}
\begin{center}
\includegraphics[width=\columnwidth]{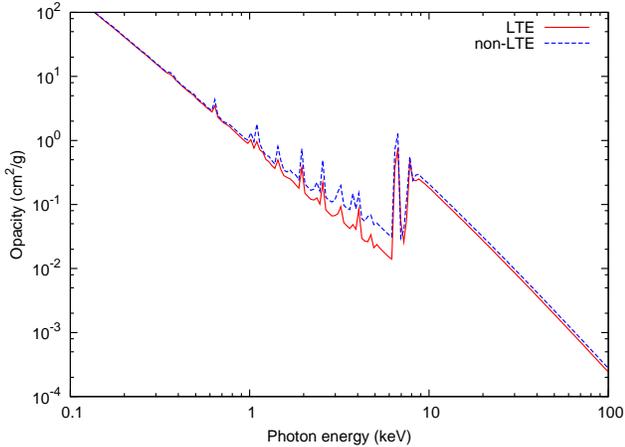}
\caption{Multigroup absorption opacity as a function of frequency
  group, for $\rho = 10^2$~\cmsqperg and $k_{\rm B}T = 10$~keV. Both
  LTE and non-LTE opacities are shown; the non-LTE opacities are
  generated using the RADIOM model (see text).}
\label{fig:RADIOM}
\end{center}
\end{figure}

The monochromatic opacities contained within the OPLIB database are
generated under the assumption that the material is in LTE, i.e., that
the bound-electron populations and the free-electron energy
distribution can be accurately described by Saha-Boltzmann
statistics. By using the database we implicitly adopt this assumption
in our paper; this also allows us to derive the emission coefficients
from the database results, using Kirchoff's law of thermal radiation
[Equation~(\ref{eq:kirchoff})]. At the base of the XRB atmosphere the
radiation-material system is in thermal equilibrium
(Section~\ref{sub:radmat}). However, at optical depths of a few or
less, this is no longer true: the radiation field is highly
non-Planckian and $T \ne T_r$. Note that the material can be in LTE
even when it is not in equilibrium with the radiation field
\citep[i.e., when $T \ne T_r$;
  e.g.,][]{rybickilightman1986,castor2004}. This happens when
material-material collisions dominate radiation-material interactions
(see Section~\ref{sub:plasma}). As is discussed in
Section~\ref{sub:plasma}, in the outer layers of the atmosphere the
material is not in thermal equilibrium even with itself. Assuming LTE
conditions in these layers can introduce large errors in the
bound-electron contributions to the frequency-dependent absorption
opacities there. However, the error in the outgoing spectrum should be
much less, due to two mitigating factors: first, the layers where LTE
conditions break down are at low optical depths ($\tau^{\rm tot} \alt
0.1$), such that they contribute very little to the spectrum; and
second, at the high temperatures and mostly hydrogen compositions of
the XRB atmosphere (but see Section~\ref{sub:mixing}) bound
transitions contribute only a small part of the total opacity. We did
not implement non-LTE opacities in our model, so we can not compare
LTE and non-LTE outgoing spectra; however, in Figure~\ref{fig:RADIOM}
we show for typical conditions in the outer layers the
frequency-dependent LTE absorption opacities and their non-LTE
approximations generated using the \mbox{RADIOM} model \citep[see
  Section~\ref{sub:plasma};][]{busquet93}. For certain frequencies the
difference in the absorption opacity is a factor of two or more.

For our scattering opacity we use the Klein-Nishina differential cross
section \cite[e.g.,][]{rybickilightman1986}, which is exact for
Compton scattering of unpolarized radiation. Using the exact
Klein-Nishina cross section rather than the Thomson approximation
(which is $\kappa_\nu^{\rm sc} \simeq Y_e \sigma_{\rm Th}/m_u$) is
essential for accurate modeling of high-luminosity atmospheres. In
particular, since the Klein-Nishina total cross section is smaller
than that of the Thomson approximation at high frequencies, we can
have $l_{\rm proj} > 1$, i.e., a projected luminosity greater than the
Thomson Eddington luminosity of Equation~(\ref{eq:LTh}), without mass
loss. In addition, due to the drop in the Klein-Nishina cross section
at high frequencies, the frequency-integrated scattering opacity
decreases with increasing temperature at such a rate that $F < F_{\rm
  crit}$ [Equation~(\ref{eq:Fcritcond})] throughout the atmosphere:
even though $r^2F$ increases toward the base of the atmosphere due to
the strong general relativistic effects there, $r^2F_{\rm crit}$
increases faster \citep[see
  Section~\ref{sec:method};][]{paczynski86,spw12}.

For $n_e$ in our scattering cross section formula
[Equation~(\ref{eq:kappascat})] we use the number density of all
electrons, bound or free. This is because, for typical photons in the
atmosphere, the energy $h\nu \agt 100$~eV is much larger than the
binding energy of any bound electrons, such that the photons see all
electrons as effectively free. In this regime the scattering cross
section for bound electrons, like that for free electrons, is given by
the Klein-Nishina form; Rayleigh scattering and other forms of
bound-electron scattering can be ignored
\citep{eisenberger70,rybickilightman1986}.

\begin{figure}
\begin{center}
\includegraphics[width=\columnwidth]{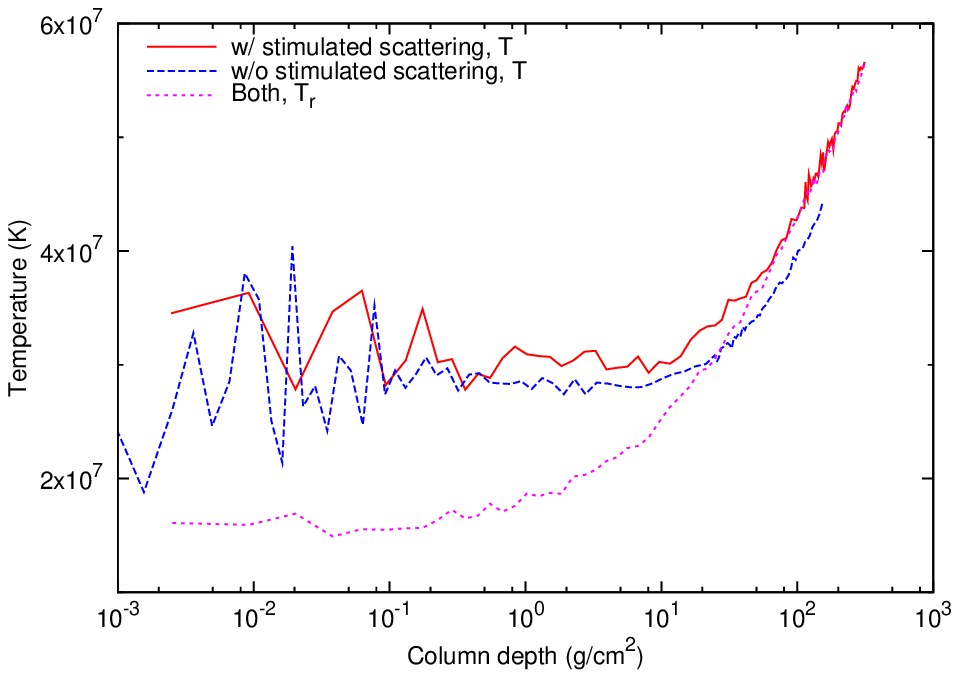}
\includegraphics[width=\columnwidth]{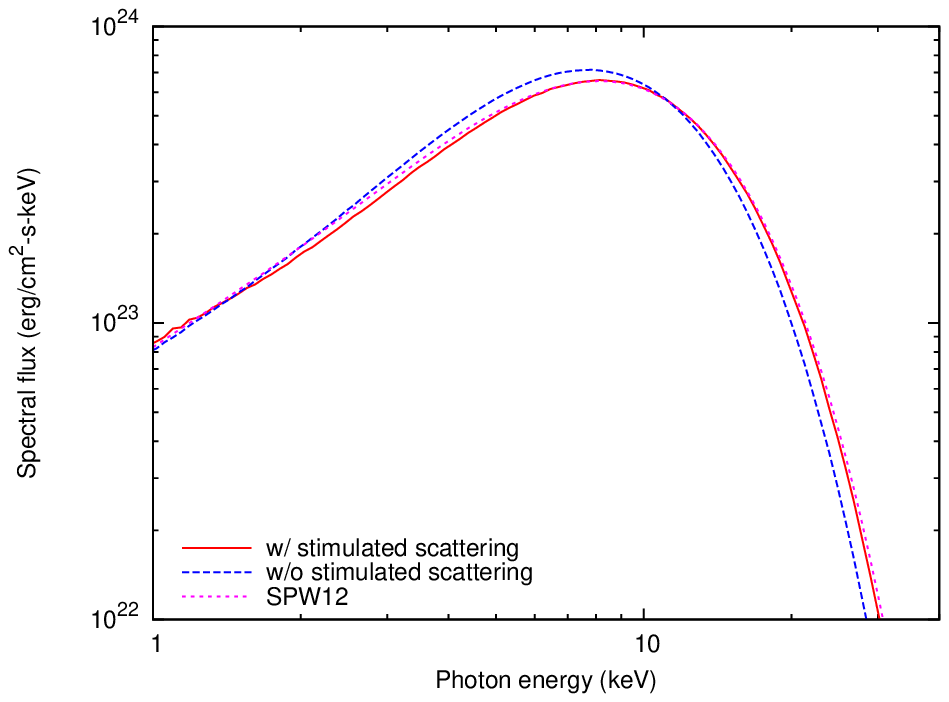}
\caption{Material and radiation temperature profiles (top panel) and
  the outgoing radiation spectrum (bottom panel), for a
  solar-composition, $g_{\rm base} = 10^{14}~{\rm cm~s^{-2}}$, $l_{\rm
    proj} = 0.98$ model atmosphere with and without stimulated
  scattering.}
\label{fig:stimulatedSpec}
\end{center}
\end{figure}

To simulate repeated scatterings of photons by a distribution of
electrons we solve the Boltzmann equation for Compton scattering; our
method is discussed in the Appendix. \citet{spw12} compared XRB
atmosphere models solving the full Boltzmann equation with ones
solving the Kompaneets approximation \citep[see,
  e.g.,][]{rybickilightman1986}, and found differences of around $5\%$
in the outgoing spectra at near-Eddington luminosities. Note that
while the direct scattering terms in the Boltzmann equation are
naturally handled by the Monte Carlo method, the stimulated scattering
terms [$1+n_\nu$ and $1+n_{\nu'}$ in Equations~(\ref{eq:kappascat})
  and (\ref{eq:jscat})] are not easily accounted for. In our
calculations we have included an approximation to stimulated
scattering, described in the Appendix. Stimulated scattering is most
important at high temperatures and densities. Without it, $T \ne T_r$
in the deep parts of the atmosphere. In the hottest atmospheres, many
high-frequency photons originate from layers where stimulated
scattering is important, such that there is a noticeable change in
both the outgoing spectra and the color correction factors when
stimulated scattering is included in our models compared with when it
is ignored. Figure~\ref{fig:stimulatedSpec} shows the effect of
stimulated scattering on a hot atmosphere, both for the temperature
profile and the outgoing spectrum.

\subsection{Plasma physics}
\label{sub:plasma}

In the neutron star atmosphere, ions and free electrons collide much
more frequently with each other than with photons. For example,
electron-electron collision rates are around $10^{14}~(\rho/1~{\rm
  g~cm^{-3}})~{\rm s^{-1}}$ \citep[e.g.,][]{NRL}, while
electron-photon rates are five orders of magnitude lower. We therefore
assume that the ions and (free) electrons are in equilibrium with
themselves and each other, i.e., that they are Maxwellian and have the
same temperature $T_i = T_e$. We assume that the ions and electrons
have Maxwell-Boltzmann, i.e., non-relativistic, distributions. This is
an excellent assumption for the ions, since $k_{\rm B}T \ll m_p c^2$,
but leads to an error of $\alt 2\%$ in the cumulative distribution for
the electrons at the highest temperatures considered here. Since the
XRB atmosphere temperatures are large ($\agt 1$~keV), the ions and
electrons are also assumed to be ideal gases.

Note that while the free electrons, which are in equilibrium, obey
Maxwell-Boltzmann statistics, the bound electrons do not
necessarily. This is because the atomic level populations are set by
the balance between radiative and {\it inelastic} collisions, which
transfer energy to or from the electrons and cause them to transition
between levels; the great majority of electron and ion collisions are
elastic and do not affect this balance
\citep[e.g.,][]{mihalasmihalas1984}. Therefore, while it is
appropriate to use LTE-derived opacities/coefficients for electron
scattering and free-free absorption/emission throughout the
atmosphere, it may not be for bound-free or bound-bound electron
contributions (see Section~\ref{sub:opacity}). The point at which LTE
conditions begin to break down for bound electrons can be estimated
from the RADIOM model \citep{busquet93,busquet09} using the parameter
\be
\beta_{\rm RAD} \sim 0.3 (T_e / 1~{\rm keV})^{7/2} (\rho / 1~{\rm g~cc^{-1}})^{-1} \,.
\ee
When $\beta_{\rm RAD}$ is larger than unity, the radiative rates
dominate the collisional rates for transitions between neighboring
ionization levels. For example, at $T_e = 1$~keV the XRB atmosphere is
in LTE for $\rho \agt 0.3~{\rm g~cc^{-1}}$. According to the RADIOM
model, the bound-free and bound-bound contributions to the opacity at
a temperature $T_e$ can be estimated by their LTE values at a
temperature
\be
T_z \simeq \frac{T_e}{(1+4\beta_{\rm RAD})^{0.2}};
\ee
this is the approximation we used to generate Figure~\ref{fig:RADIOM}.

\subsection{Hydrodynamics}
\label{sub:hydro}

For simplicity, we do not solve the full radiation-hydrodynamics
coupled equations, but use an iterative hydrostatic method
\citep[Section~\ref{sub:hydrosolve}; cf.][]{ebisuzaki1987}. That this
is a reasonable approximation for our models can be seen by examining
the gas sound speed, given for an ideal gas by $c_s = \sqrt{\gamma
  P_g/\rho} = \sqrt{\gamma k_{\rm B}T/m_p} \sim 10^{-3}c$. Even at the
base of the atmosphere this is an order of magnitude smaller than the
radiation diffusion velocity (Section~\ref{sub:radmat}), such that the
time scale for the adjustment of the atmosphere structure is at least
an order of magnitude larger than the time scale for the adjustment of
the radiation field. Note that even for the thickest atmospheres we
model here, a few~km thick, the sound crossing time ($\sim 0.01$~s) is
less than the typical time scale for changes in the X-ray burst
\citep[$0.1$--$1$~s; see][]{gallowayetal2008}. As long as the
atmosphere remains in this thickness regime, it will evolve from one
quasi-static state to the next as the burst grows or decays. In the
future, if we extend our work to thicker atmospheres, with sound
crossing times comparable to the X-ray burst rise time
(Section~\ref{sec:discuss}), we will have to model the hydrodynamic
processes more accurately.

\subsection{Gravity}
\label{sub:gravity}

We consider general relativistic effects by solving the radiation
transfer equation in a Schwarzschild geometry. Such a complication is
required for models of extended atmospheres, since without the
$(1+z)^{-2}$ scaling of $r^2F$ with radius provided by the
Schwarzschild metric these atmospheres would be hydrodynamically
unstable [Section~\ref{sub:initial};
  cf.\ Equation~(\ref{eq:kpcrit})]. However, it is not strictly
necessary for models of thin atmospheres $r-r_{\rm base} \equiv \Delta
r \alt 10^5$~cm (such as those considered in
Section~\ref{sec:results}), since relativistic, thin atmospheres are
almost identical to their Newtonian counterparts
\citep{madej04,spw11,spw12}. This is because general relativistic
effects depend on the change in the metric, which is of order $\Delta
r/r \alt 10^{-1}$ across these atmospheres; or on the integrated
radial and angular deviations in the case of light bending, which are
both of order $(2GM/c^2r) (\Delta r/r) \alt 10^{-2}$ (see below).

We treat light bending as a perturbation on the photon transport,
calculating the transport distance $r_{\rm new}-r_{\rm old}$ in the
Newtonian limit and then using this distance to calculate the
general-relativistic change in direction. From
Equation~(\ref{eq:GRtraj}), we have to first order in $2GM/c^2r$ that
the deviation of $dr/d\theta$ from the straight-line trajectory of a
photon in free space ($2GM/c^2r \rightarrow 0$) is
\be
\frac{dr_{\rm dev}}{d\theta} = -\frac{GM}{c^2b}\frac{b^2}{\sqrt{r^2-b^2}}
\label{eq:rdev}
\ee
such that
\be
\frac{dr_{\rm dev}}{dr} = \frac{GM}{c^2r}\frac{b^2}{r^2-b^2} \,,
\ee
while the deviation of $d\theta/dr$ is
\be
\frac{d\theta_{\rm dev}}{dr} = -\frac{GM}{c^2r}\frac{b^3}{r\left(r^2-b^2\right)^{3/2}} \,.
\label{eq:phidev}
\ee
For most transport directions, we have $b^2/(r^2-b^2) \alt 1$ [and for
  directions where this is not true,
  Equations~(\ref{eq:rdev})--(\ref{eq:phidev}) are not valid
  anyway]. Therefore, our approximation introduces both distance- and
direction-related errors of order $(2GM/c^2r) (\Delta r/r)$.

Our models have three free parameters, $\{X\}$, $g_{\rm base}$, and
$l_{\rm proj}$, that we vary to generate a series of atmospheres; the
total and spectral fluxes from these atmospheres can be fit to
observations to constrain the mass and radius of a given neutron star
or set of neutron stars (Sections~\ref{sec:results} and
\ref{sec:discuss}). Ideally we should also vary $r_{\rm base}$, to
generate a larger series of atmospheres. However, as we discussed
above, changing $r_{\rm base}$ by tens of percent makes no difference
for the majority of our atmospheres, which have thicknesses much less
than the stellar radius; these thin atmospheres only depend on
composition, surface gravity, and luminosity. Therefore, we instead
fix $r_{\rm base} = 11.5$~km, a typical neutron star radius from the
models of \citet{steiner10,steiner13}, with the intent that this
$r_{\rm base}$ is only to be used as an order-of-magnitude estimate
for generating atmospheres and is not to be taken as the actual radius
of an observed neutron star. Since changing $r_{\rm base}$ does make a
difference in extended atmospheres, in future work we will vary this
parameter in our models for more accurate fits to X-ray bursts with
strong atmosphere expansion.

In our models, we can safely ignore the atmosphere mass, pressure, and
energy density when calculating the strength of various general
relativistic effects: The column depth at the base of the atmosphere
is around $100~{\rm g~cm^{-2}}$, such that the atmosphere mass $\sim
10^{36}~{\rm erg}/c^2$ is much less than the total mass of the neutron
star $M \sim 10^{54}~{\rm erg}/c^2$. Similarly, the contribution of
the local pressure to the gravitational acceleration $4\pi r^3P \alt
10^{40}~\erg$ for $T_r \sim 10~\keV$ (where the upper limit is for
extended atmospheres with thicknesses of order tens~of~km) is much
less than the contribution of the neutron star mass ($Mc^2$). The
assumption $P_{rr} \ll \rho c^2$ breaks down for $\rho \alt
10^{-7}~{\rm g~cm^{-3}}$ (assuming $T_r \sim 1~\keV$ in the outer
atmosphere); however, even in extended atmospheres this region
(optical depth $\alt 0.1$) will contribute very little to the outgoing
radiation.

\subsection{Compositional mixing}
\label{sub:mixing}

On its own, the strong gravity on a neutron star would quickly
\citep[on the order of seconds; e.g.,][]{lai01} separate the outer
layers by chemical species, such that the atmosphere would be composed
entirely of the lightest species, hydrogen. However, this separation
is counteracted by several processes: First, diffusion between layers
of different species ensures that the atmosphere composition will not
be uniform. In the atmosphere, the compositional gradient due to the
balance between gravity and diffusion is most likely small
\citep[e.g.,][]{chang.bildsten:diffusive}, but its effect on the
outgoing spectrum should be investigated. Second, accretion provides
new material to the top of the atmosphere that is not immediately
separated by gravity. If the accretion rate is large enough, it will
modify the atmosphere composition in an observable way. Third, mass
loss during PRE can change the atmosphere composition by exposing the
underlying layers
\citep{ebisuzaki1987,ebisuzakinakamura1988}. Convection by itself can
not affect the atmosphere composition, since the convective mixing
zone does not reach the base of the atmosphere. Instead, convection
mixes heavy elements (the ashes of the thermonuclear burning powering
the XRB) to just below the base, and if PRE mass loss is large enough
these heavy-element-enriched layers will become the new atmosphere
\citep{weinbergetal2006,kajava16}.

An accurate consideration of the effects of compositional mixing on
the atmosphere is difficult. The composition of the accreted material
is typically unknown (\citealt{strohmayerbildsten2006}; but see
\citealt{gallowayetal2004}). In addition, modeling the XRB nuclear
reaction network and convection zone physics is complicated
\citep[e.g.,][]{woosleyetal2004,malone14} and outside of the scope of
this work. For simplicity, we ignore mixing effects and instead run
simulations for a variety of possible XRB atmosphere compositions (see
Section~\ref{sec:results}); ideally, these simulations will bound the
space of possible mass-radius constraints. In future work we would
like to consider mixing processes in more detail.
 
\section{Results}
\label{sec:results}

Using the method described in Section~\ref{sec:method}, we have
calculated a series of hot neutron star atmospheres. As mentioned in
that section, our models have one fixed parameter, $r_{\rm base} =
11.5$~km, and three free parameters, $\{X\}$, $g_{\rm base}$, and
$l_{\rm proj}$. For ease of comparison, we vary our free parameters
over the same range as does \citet{spw11}, though unlike in that work
we do not generate atmospheres for all possible combinations of the
parameter values. We use six compositions: pure hydrogen, pure helium,
a solar mixture $Z_\odot$, and ``fractions'' of that solar mixture
0.3$Z_\odot$, 0.1$Z_\odot$, and 0.01$Z_\odot$. Here the solar mixture
is the fifteen most-abundant elements in table~1 of \citet{asplund09}
(H, He, C, N, O, Ne, Na, Mg, Al, Si, S, Ar, Ca, Fe, and Ni) with the
number fractions calculated by normalizing the relative abundances in
that table; while the designation ``$f_Z Z_\odot$'' for a mixture
means that the mass fraction of all elements $Z > 2$ is $f_Z$ times
the corresponding mass fraction in the solar mixture, the hydrogen
mass fraction is fixed at $X_{\rm H} = 0.7374$, and the helium mass
fraction represents the remainder (such that $\sum_{i=1}^{15} X_i =
1$). We use three gravities: $\log(g_{\rm base}/{\rm cm~s^{-2}}) =
14.0$, $14.3$, and $14.6$. In addition, we use the twenty luminosity
ratios from \citet{spw11}: $l_{\rm proj}$ = 0.001, 0.003, 0.01, 0.03,
0.05, 0.07, 0.1, 0.15, 0.2, 0.3, 0.4, 0.5, 0.6, 0.7, 0.75, 0.8, 0.85,
0.9, 0.95, and 0.98; along with three others: $l_{\rm proj}$ = 1.02,
1.06, and 1.09 \citep[cf.][]{spw12}. Note that the atmospheres shown
in this section are relatively thin, with $r_{\rm surf} \alt
10^5$~cm. Our goal in this paper was to compare with XRB model results
from other groups, particularly Madej et al.\ and Suleimanov et
al.\ (Section~\ref{sec:intro}). In future work we will consider
atmospheres with $l_{\rm proj} > 1.09$ and $r_{\rm surf} > 10^5$~cm
(see Section~\ref{sec:discuss}).

For consistency, the spectral fluxes shown in this section are plotted
in terms of
\be
F_{\nu,\rm proj} = F_{\nu,\rm surf} \frac{{\cal R}_{\rm surf}^2r_{\rm surf}^2}{{\cal R}_{\rm base}^2r_{\rm base}^2}
\label{eq:Fproj}
\ee
versus
\be
\nu_{\rm proj} = \nu_{\rm surf}\frac{{\cal R}_{\rm surf}}{{\cal R}_{\rm base}} \,;
\label{eq:nuproj}
\ee
i.e., the spectra are projected on to the base of the atmosphere
[cf.\ Equation~(\ref{eq:Lproj})]. This modification has almost no
effect on the spectra for thin atmospheres, but for extended
atmospheres avoids some of the position-related ambiguity of general
relativity; see Section~\ref{sub:model}. Note that to convert to
spectral fluxes as seen at infinity (for comparison with
observations), one would use
\be
F_{\nu,\infty} = F_{\nu,\rm proj} \frac{{\cal R}_{\rm base}^2r_{\rm base}^2}{D^2} \equiv F_{\nu,\rm proj} \left[\frac{r_{\rm base}}{D (1+z_{\rm base})}\right]^2
\ee
and
\be
\nu_\infty = \nu_{\rm base} (1+z_{\rm base})^{-1} \,,
\ee
where $D$ is the distance to the neutron star.

\begin{figure}
\begin{center}
\includegraphics[width=\columnwidth]{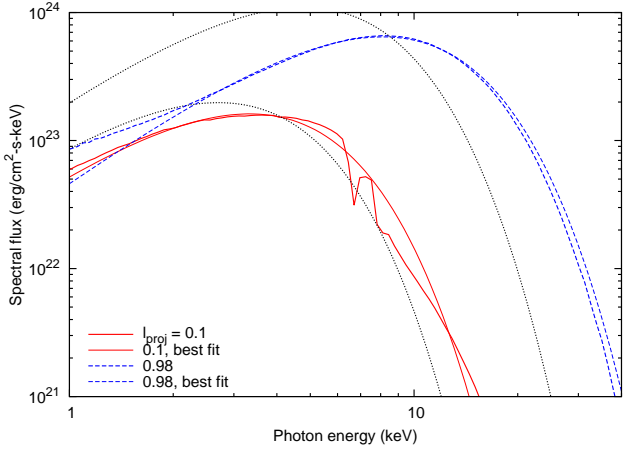}
\caption{Outgoing radiation spectrum, for atmospheres with solar
  composition, $g_{\rm base} = 10^{14}~{\rm cm~s^{-2}}$, and $l_{\rm
    proj} = 0.1$ or $0.98$. The best fit for each spectrum is shown as
  a thin line of the same color/pattern. The blackbody spectrum
  $B_\nu(T_{\rm eff,proj})$ for each atmosphere is also shown for
  comparison, as a black dotted line.}
\label{fig:modelfit}
\end{center}
\end{figure}

In this section we also plot the projected color correction (or
spectral hardness) factor
\be
f_{\rm c,proj} = \frac{T_{\rm c,proj}}{T_{\rm eff,proj}} \,,
\ee
where $T_{\rm c,proj}$ is the color temperature of the spectrum
projected on to the base of the atmosphere [using
  Equations~(\ref{eq:Fproj}) and (\ref{eq:nuproj})] and $T_{\rm
  eff,proj} = (L_{\rm proj}/4\pi r_{\rm base}^2\sigma_{\rm SB})^{1/4}$
\citep{londonetal1986,madej04,spw11,spw12}. For a given atmosphere,
the color temperature is found by fitting the spectrum to a diluted
Planck function
\be
F_{\nu,\rm proj} \simeq wB_\nu(T_{\rm c,proj}) \,;
\label{eq:diluteF}
\ee
for a perfect fit we would have $w = f_{\rm c,proj}^{-4}$. To fit our
models to Equation~(\ref{eq:diluteF}) we use the ``first'' procedure
of \citet{spw11}, which consists of varying the parameters $w$ and
$T_{\rm c,proj}$ to minimize the sum
\be
\sum_{\rm n=1}^N\left[F_{\nu_n,\rm proj} - w B_{\nu_n}(T_{\rm c,proj})\right]^2 \,,
\label{eqfit1}
\ee
where $N$ is the number of frequency groups in the (RXTE PCA) energy
band (3--20)$\times(1+z_{\rm base})$~keV [see Equation~(\ref{eq:Fnug})
  for the conversion from frequency group to frequency]. Note that
Suleimanov et al.\ treat the $1+z_{\rm base}$ factor slightly
differently than we do, and therefore fit their models over a slightly
different energy band; however, as they point out, such a change makes
a negligible difference to the color correction values
obtained. Figure~\ref{fig:modelfit} shows examples of spectra and
their best fits from our models.

\begin{figure}
\begin{center}
\includegraphics[width=\columnwidth]{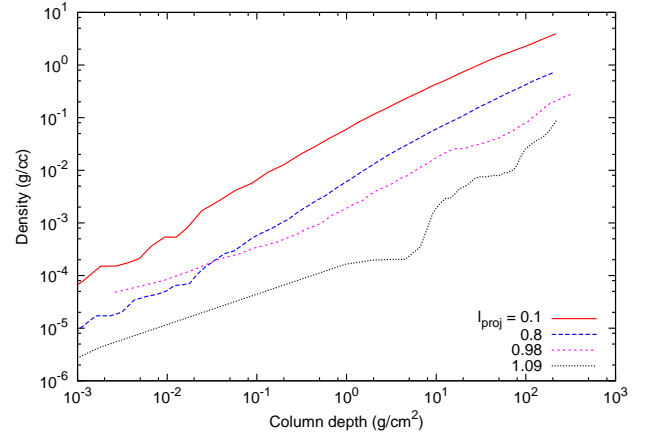}
\caption{Density as a function of column depth, for models with solar
  composition, $g_{\rm base} = 10^{14}~{\rm cm~s^{-2}}$, and a variety
  of luminosity ratios. The quantity $r-r_{\rm base}$ (see
  Figure~\ref{fig:initial}) at column depth $y = 1~{\rm g~cm^{-2}}$
  (optical depth $\tau_F^{\rm tot} \simeq 0.3$) is $130$, $850$,
  $3000$, and $4.4\times10^4$~cm for the $l_{\rm proj} = 0.1$, $0.8$,
  $0.98$, and $1.09$ model, respectively.}
\label{fig:modelrho}
\end{center}
\end{figure}

\begin{figure}
\begin{center}
\includegraphics[width=\columnwidth]{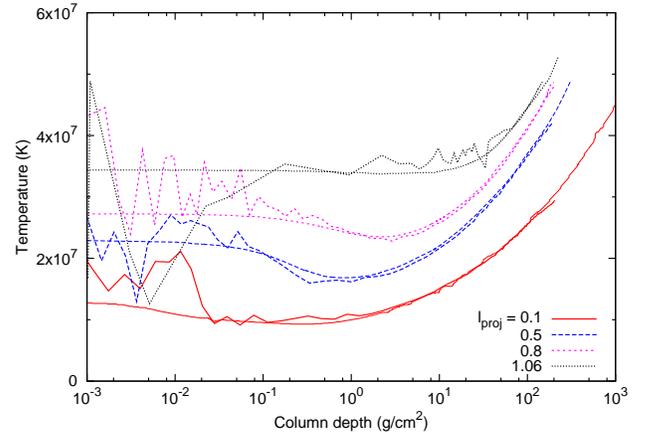}
\caption{Material temperature as a function of column depth, for
  atmospheres with solar composition, $g_{\rm base} = 10^{14}~{\rm
    cm~s^{-2}}$, and a variety of luminosity ratios. The results from
  the model of \citet{spw12} are shown as thin lines for
  comparison. The approximation $T_{\rm outer}$ from
  Equation~(\ref{eq:Touter}) is $1.05$, $2.03$, $2.52$, and $3.21$~keV
  for the $l_{\rm proj} = 0.1$, $0.5$, $0.8$, and $1.06$ atmosphere,
  respectively.}
\label{fig:spwCompT}
\end{center}
\end{figure}

\begin{figure}
\begin{center}
\includegraphics[width=\columnwidth]{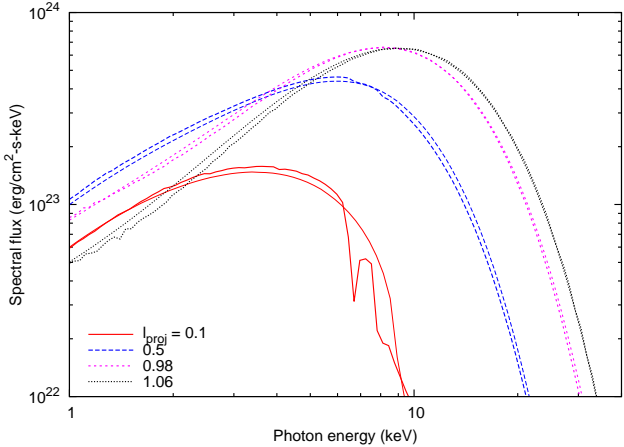}
\caption{Outgoing radiation spectrum, for atmospheres with solar
  composition, $g_{\rm base} = 10^{14}~{\rm cm~s^{-2}}$, and a variety
  of luminosity ratios. The results from the model of \citet{spw12}
  are shown for comparison, as thin lines with the same
  colors/patterns as our results.}
\label{fig:spwCompSpecHigh}
\end{center}
\end{figure}

Figures~\ref{fig:modelrho} and \ref{fig:spwCompT} show, respectively,
examples of the density and temperature as a function of column depth
from our models; while Figure~\ref{fig:spwCompSpecHigh} shows an
example of the outgoing radiation spectrum. Figures~\ref{fig:spwCompT}
and \ref{fig:spwCompSpecHigh} also show the results of
\citet{spw11,spw12} for comparison (cf.\ figure~3 of either work). Our
temperature profiles are qualitatively similar to those of Suleimanov
et al.; compared to the results of \citet{madej04}, however, our
temperature profiles have a significantly larger dip at column depths
of order unity \citep[see, e.g., figure~C.1 of][]{spw12}. The profiles
from both our work and that of Suleimanov et al.\ approach $T_{\rm
  outer}$ in the outer layers to within a few percent. Similarly, our
spectra are qualitatively similar to those of Suleimanov et al.\ but
differ substantially from those of Madej et al.\ (see below).

\begin{figure*}
\begin{center}
\begin{tabular}{c c c}
\includegraphics[angle=-90,width=0.66\columnwidth]{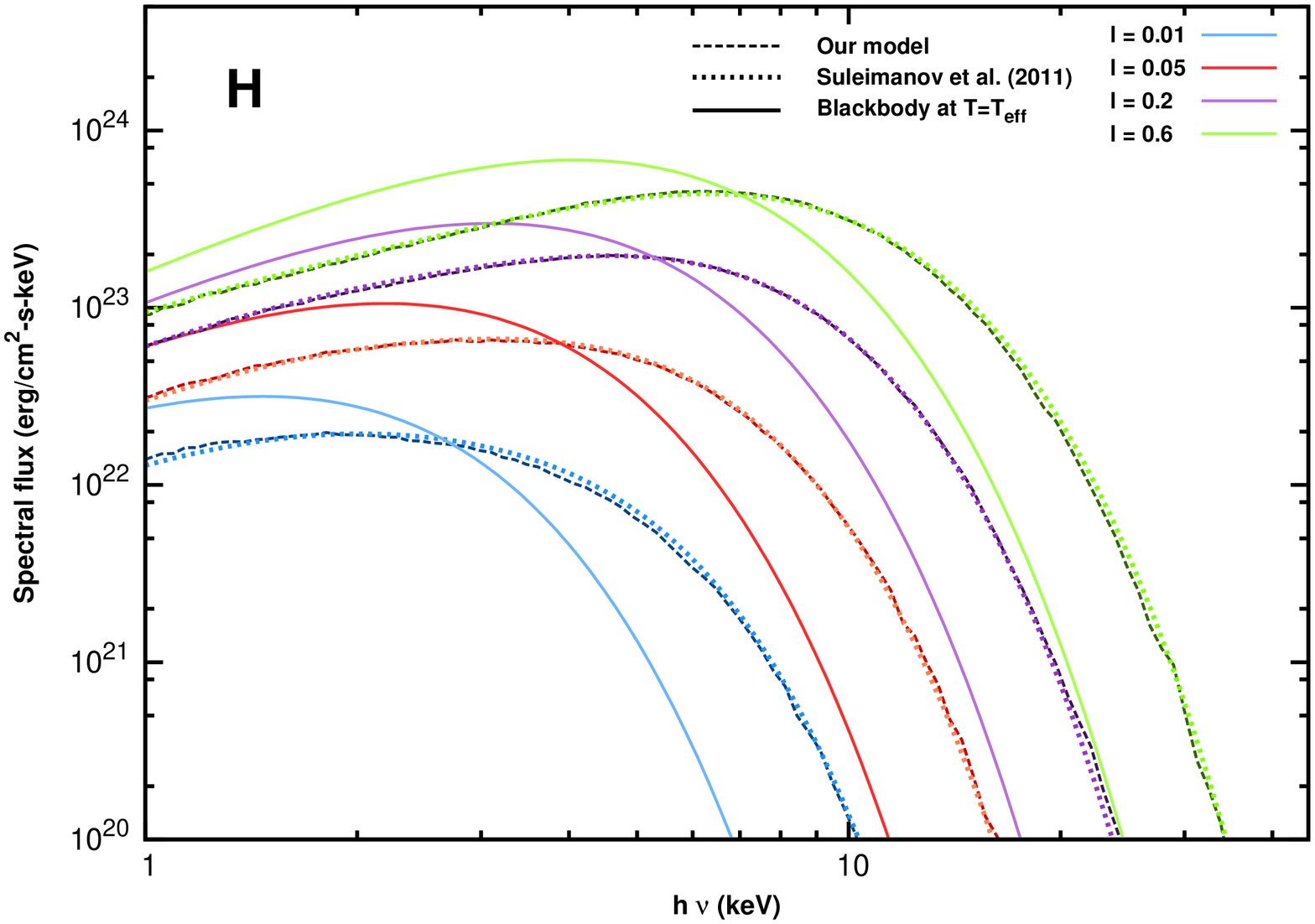} &
\includegraphics[angle=-90,width=0.66\columnwidth]{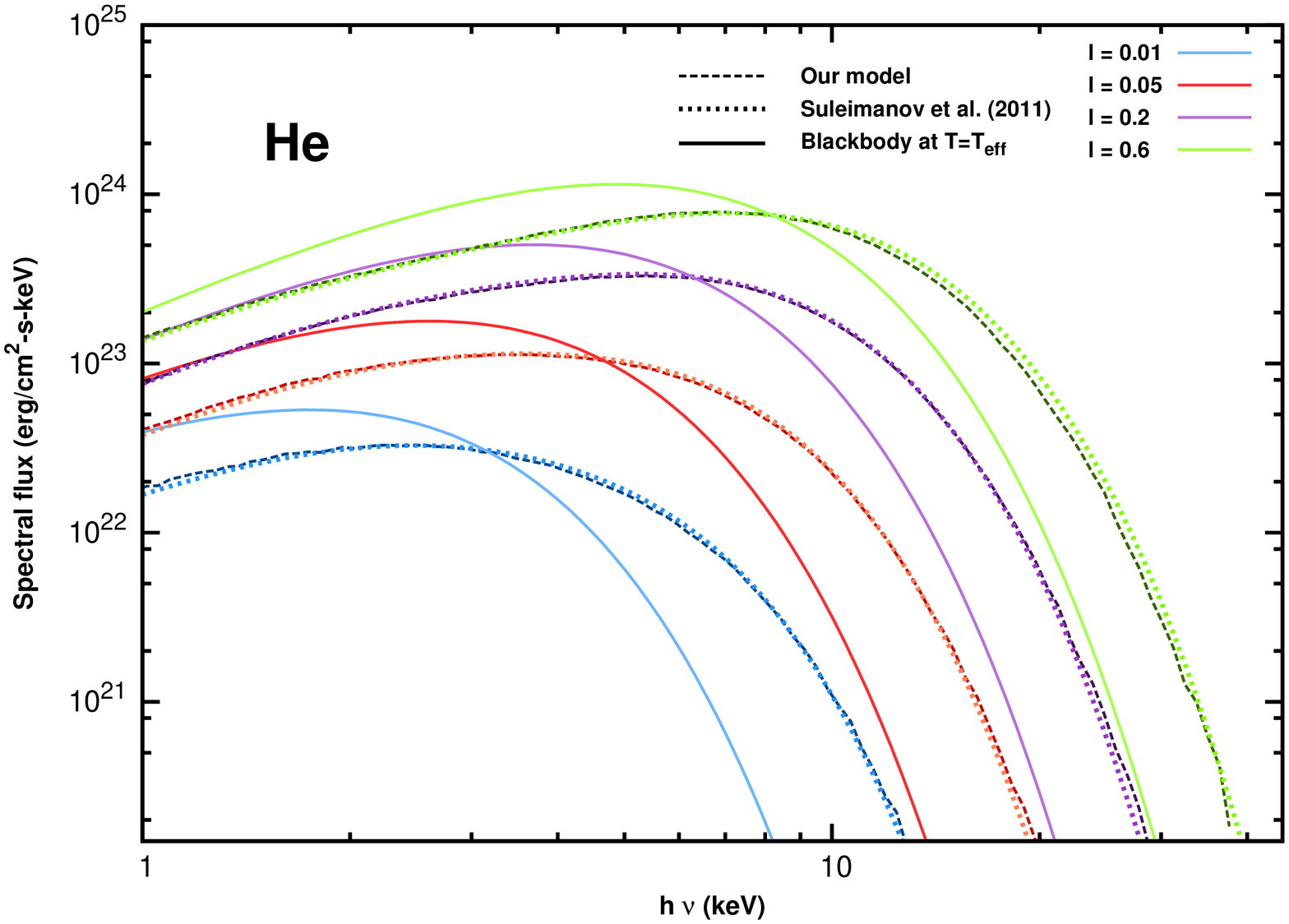} &
\includegraphics[angle=-90,width=0.66\columnwidth]{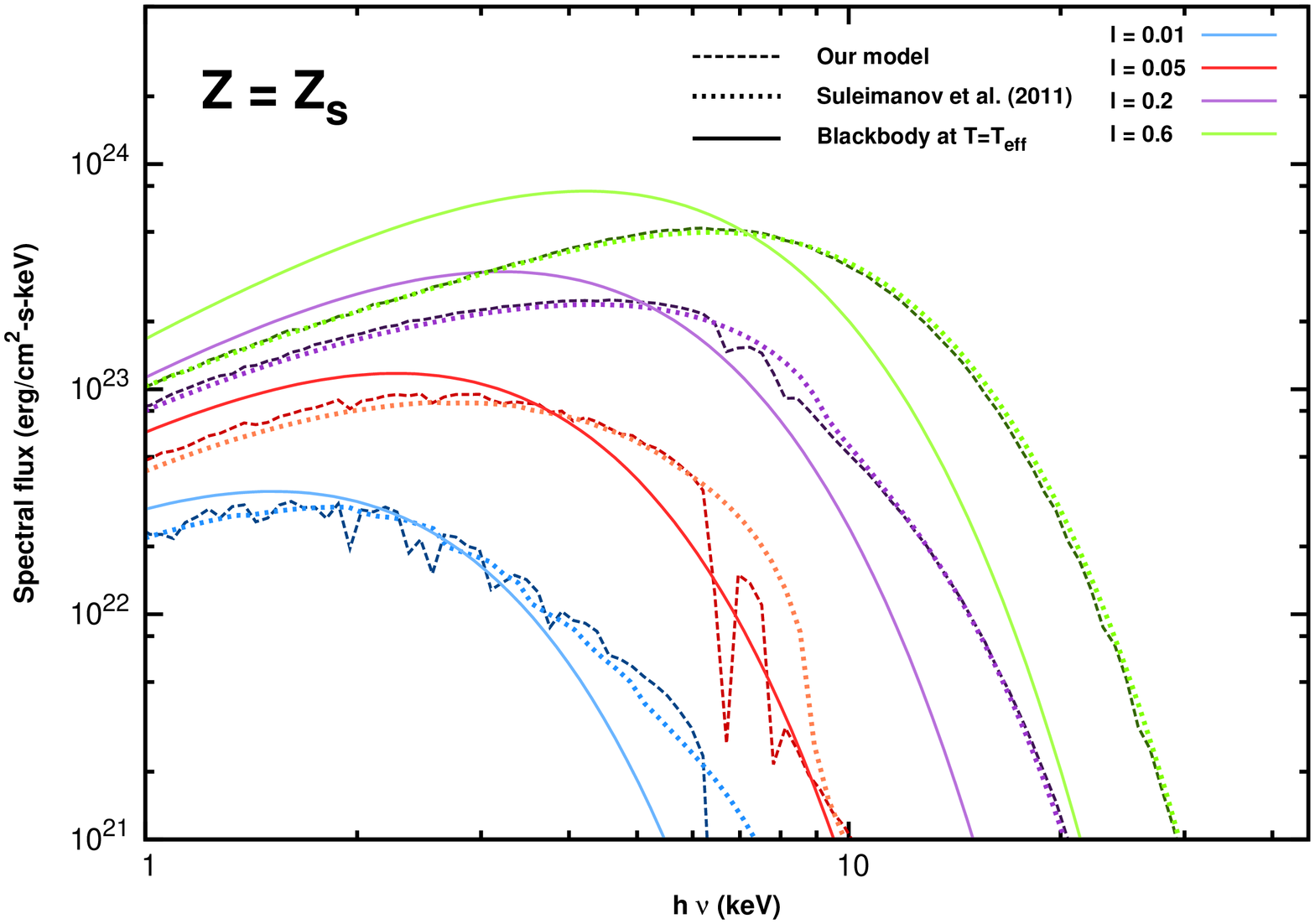} \\
\includegraphics[angle=-90,width=0.66\columnwidth]{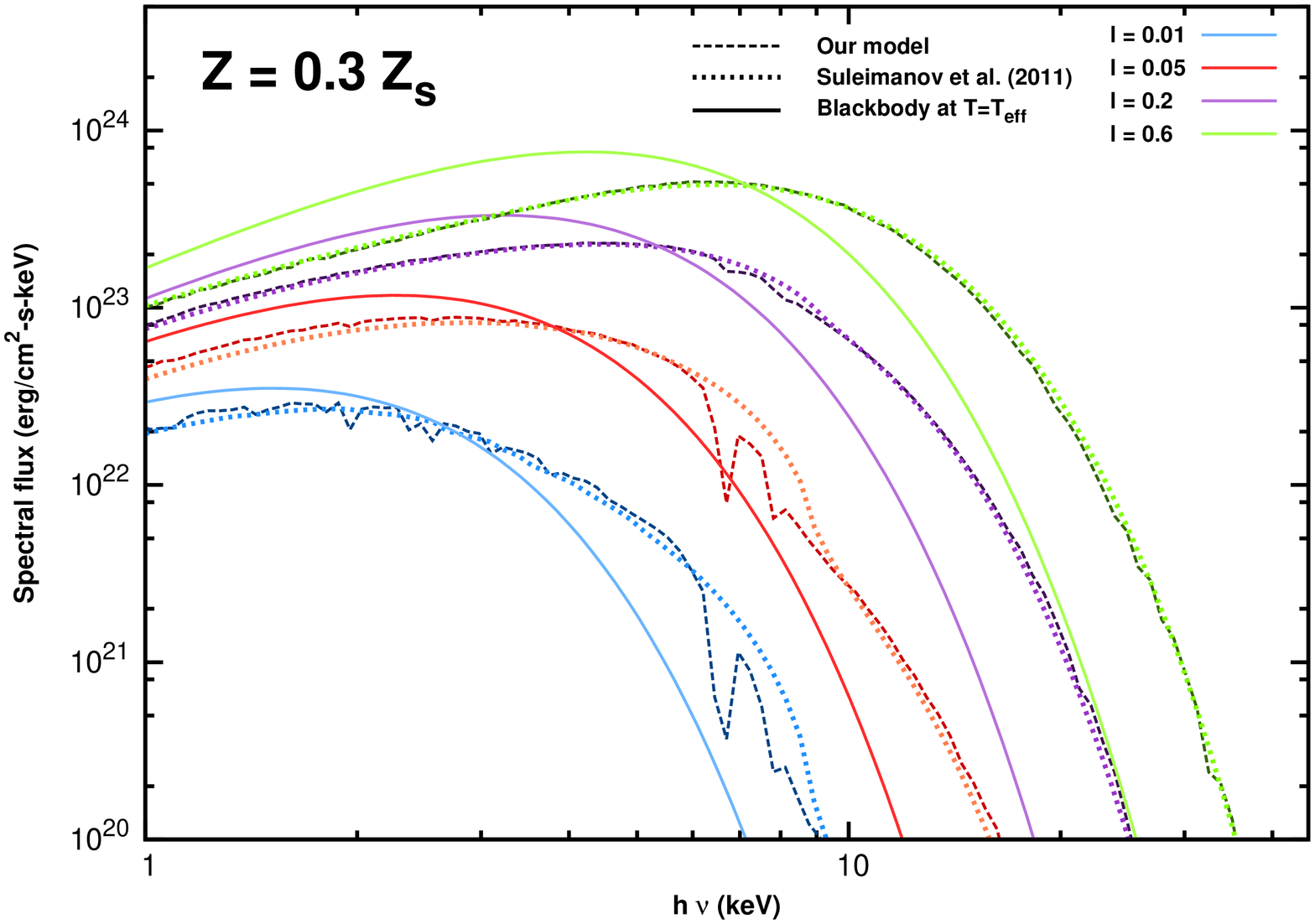} &
\includegraphics[angle=-90,width=0.66\columnwidth]{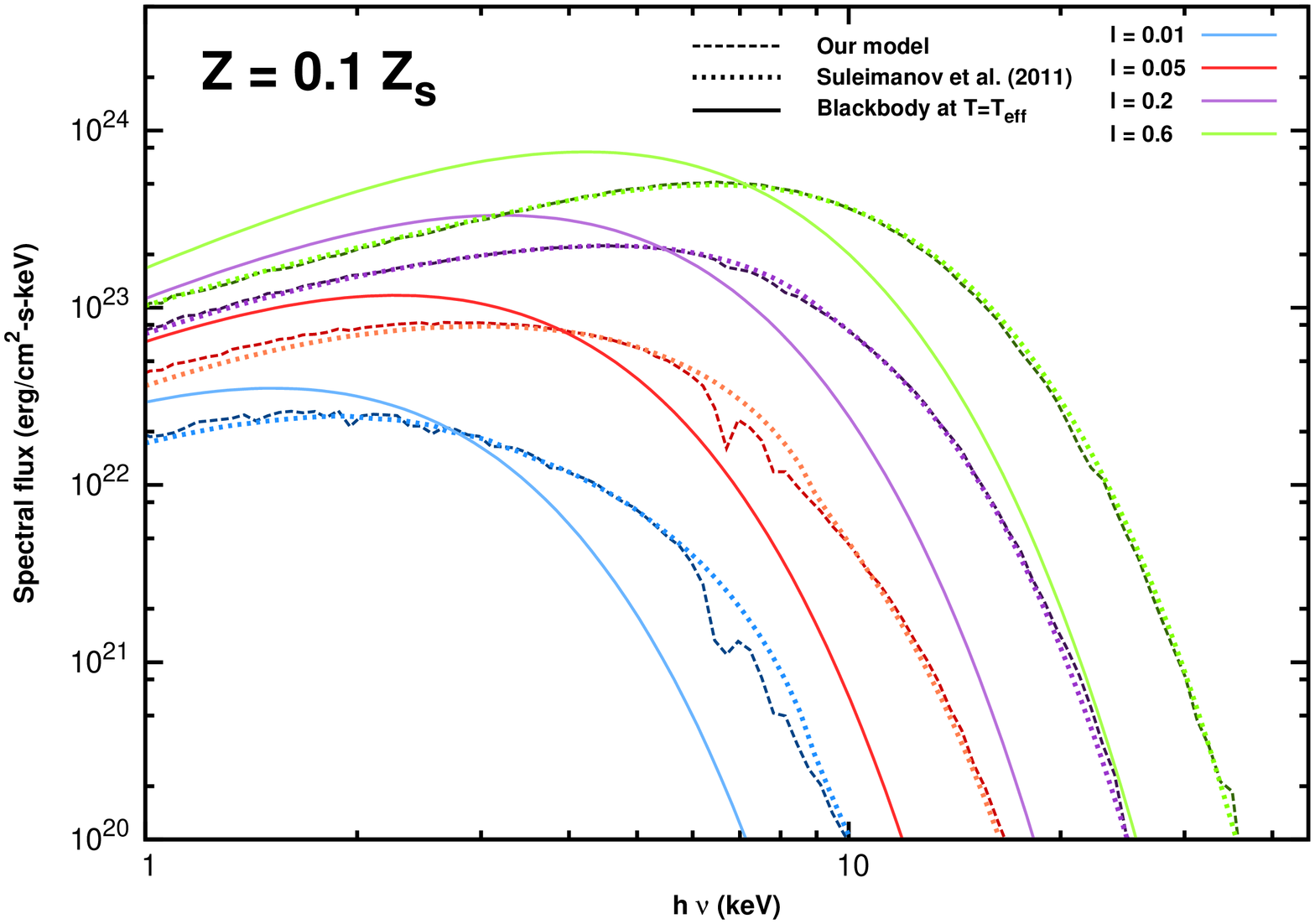} &
\includegraphics[angle=-90,width=0.66\columnwidth]{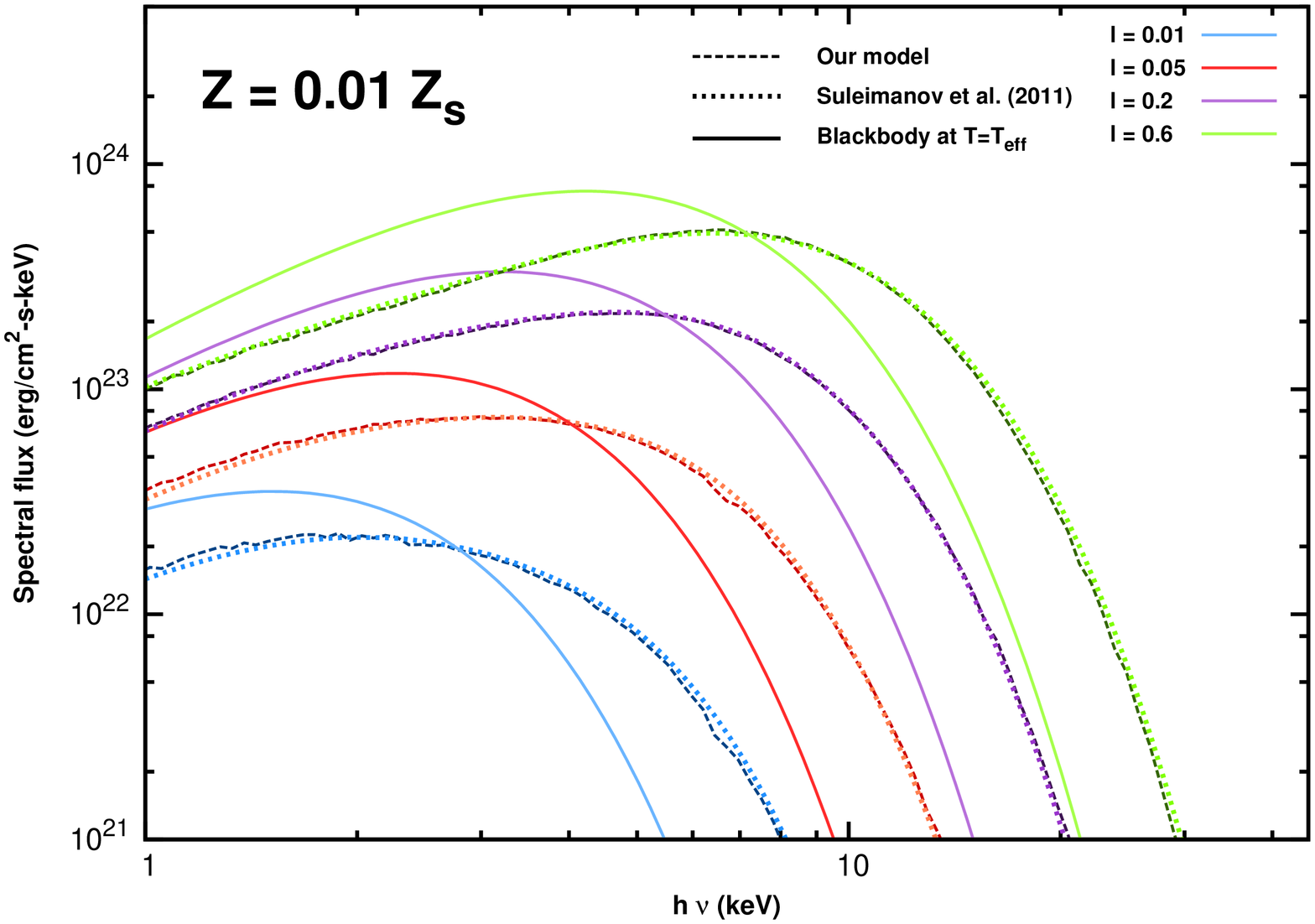}
\end{tabular}
\caption{Outgoing radiation spectrum for a variety of atmospheres
  (dashed lines). For comparison, the blackbody approximations are
  shown as solid lines and the results from the model of
  \citet{spw12} are shown as dotted lines. Here the surface gravity
  is $g_{\rm base} = 10^{14}~{\rm cm~s^{-2}}$.}
\label{fig:spwSpecComp}
\end{center}
\end{figure*}

\begin{figure}
\begin{center}
\includegraphics[angle=-90,width=\columnwidth]{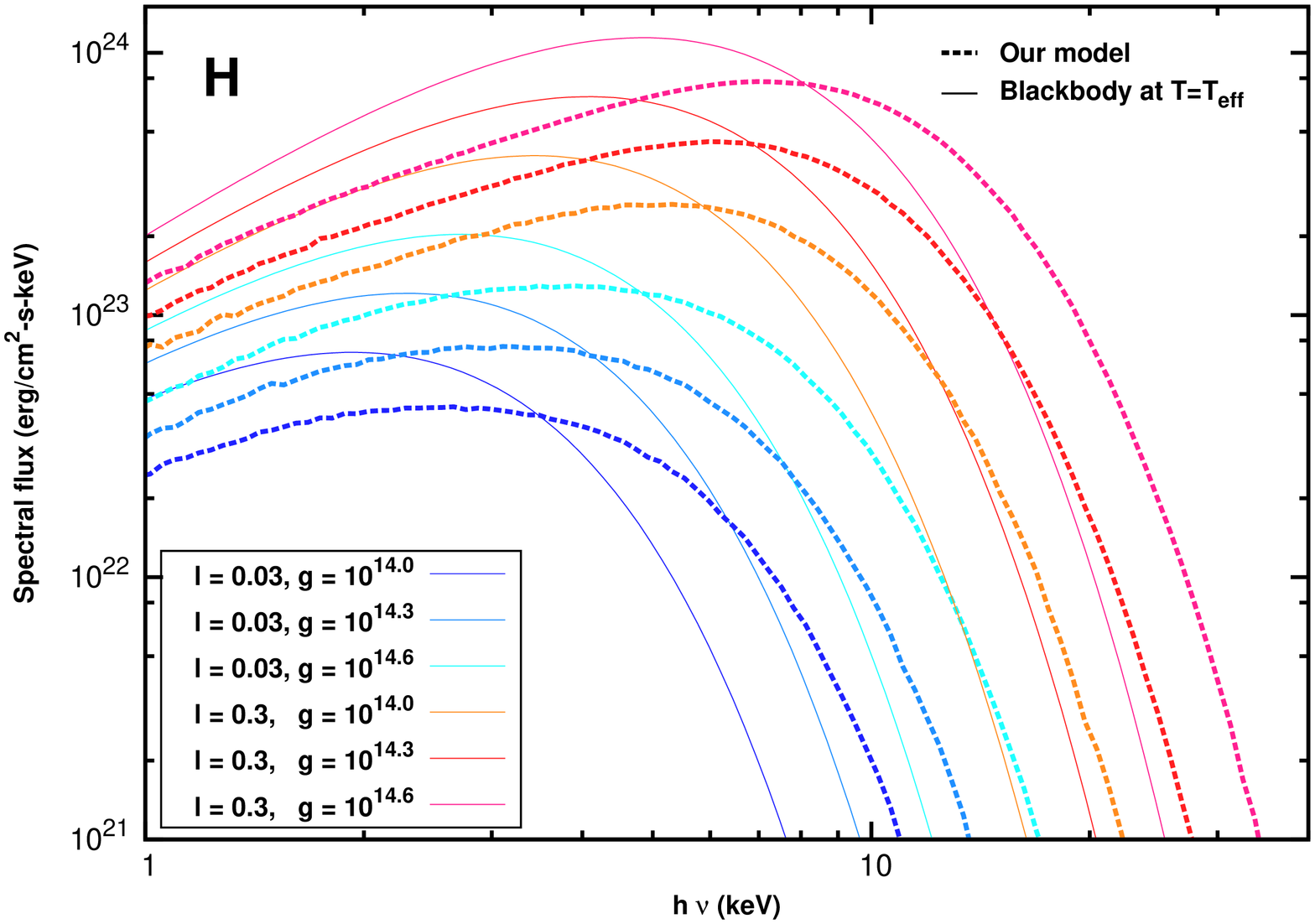}
\includegraphics[angle=-90,width=\columnwidth]{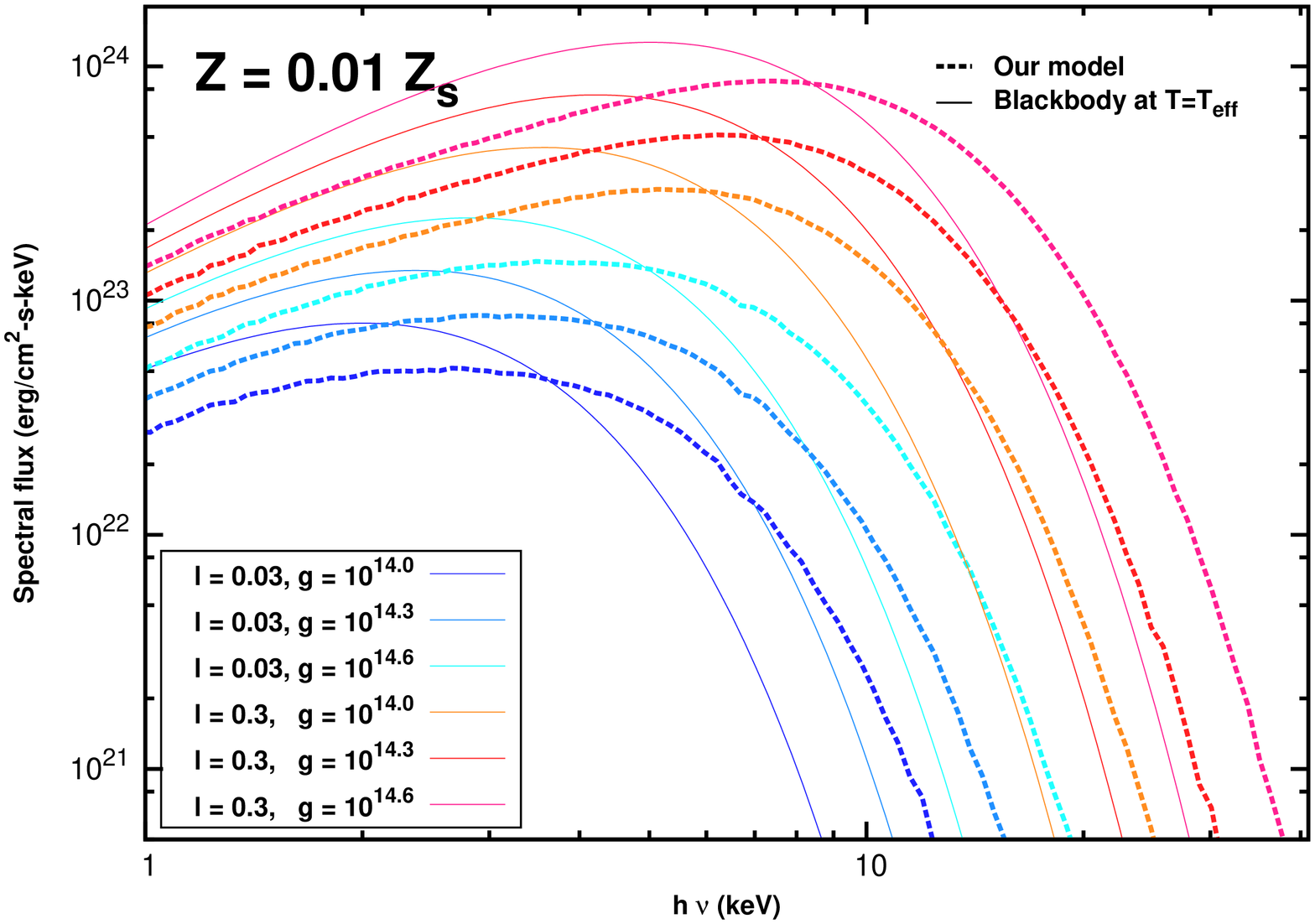}
\caption{Outgoing radiation spectrum for a variety of atmospheres
  (dashed lines). For comparison, the blackbody approximations are
  shown as solid lines. Here the composition is pure hydrogen (top
  panel) or modified solar (0.01$Z_\odot$; bottom panel).}
\label{fig:spwSpecComp2}
\end{center}
\end{figure}

Figure~\ref{fig:spwSpecComp} shows a comparison of the outgoing
spectra from our models and those of Suleimanov et al., for a variety
of atmosphere compositions and luminosity ratios and a single gravity
$\log(g_{\rm base}/{\rm cm~s^{-2}}) = 14.0$;
Figure~\ref{fig:spwSpecComp2} shows a comparison for a variety of
gravities. The spectra from the two works are very similar, except at
the lowest luminosities and highest metallicities considered. In these
low-$L$, high-$X_{\rm metal}$ cases, the spectra agree qualitatively
but differ in the number of absorption features and the amplitudes of
these features, owing to the different opacities used. As a
consequence, the color corrections derived from these spectra are also
different (see below). Conversely, the \citet{madej04} spectra have a
qualitatively different shape, including a different peak and low- and
high-frequency falloffs (again, compare our $g_{\rm base} =
10^{14.3}~{\rm cm~s^{-2}}$, $l_{\rm proj} = 0.5$ results to Madej et
al.'s $T_{\rm eff} = 2\times10^7$ results; or see figure~C.1 of
\citealt{spw12}). Note that if we had not included stimulated
scattering in our models, our spectra would fall off faster with
frequency at the high-frequency end and would not match as closely
with the results of Suleimanov et al.\ (see
Figure~\ref{fig:stimulatedSpec} in Section~\ref{sub:opacity}). On the
other hand, for low frequencies our spectra with stimulated scattering
differ by a few percent from those of Suleimanov et al., but without
stimulated scattering are nearly exact; we attribute this difference
to our approximate treatment of stimulated scattering (see the
Appendix).

\begin{figure}
\begin{center}
\includegraphics[width=\columnwidth]{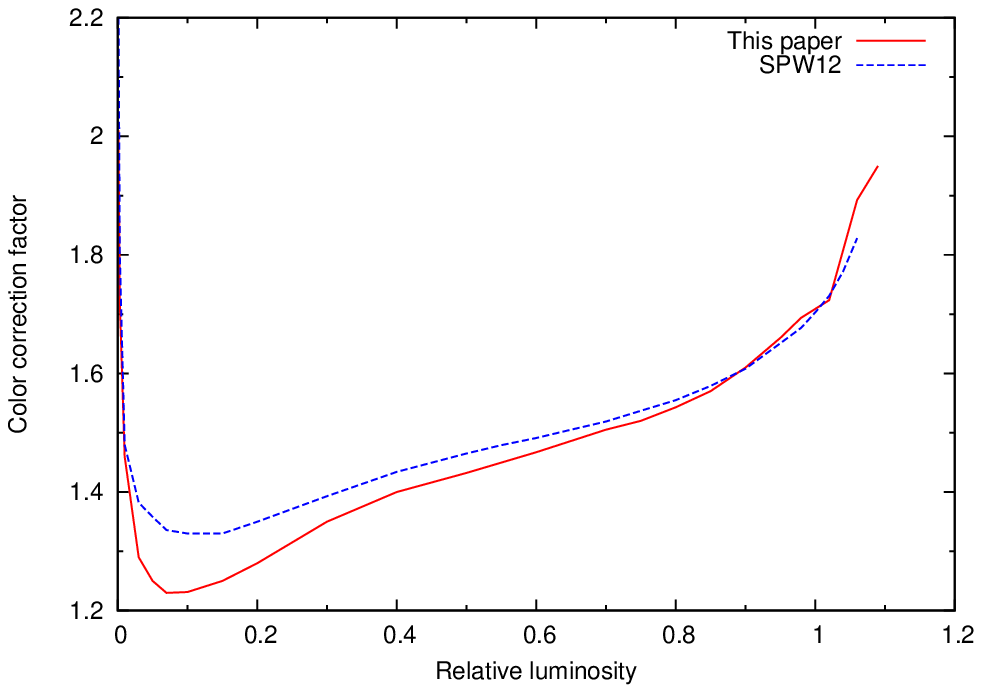}
\caption{Color correction factor as a function of relative luminosity
  for solar composition and $g_{\rm base} = 10^{14}~{\rm
    cm~s^{-2}}$. The results from the model of \citet{spw12} are shown
  as a dashed curve for comparison.}
\label{fig:fc}
\end{center}
\end{figure}

Figure~\ref{fig:fc} shows another comparison between our models and
those of Suleimanov et al., in this case the color correction factor
as a function of luminosity ratio for one atmosphere composition and
gravity. A fixed $\{X\}$ and $g_{\rm base}$ corresponds to one burst
on a hypothetical neutron star, assuming that the chemical composition
in the atmosphere remains constant during the burst
(Section~\ref{sub:mixing}). The curves from the two works are similar
at high luminosity ratios, but differ by about 10\% at low ratios
($l_{\rm proj} \alt 0.4$), due to the previously mentioned differences
in the bound-free/bound-bound opacities used by each work. The large
difference in $f_{\rm c,proj}$ for spectra that are very similar
highlights the fact that at low luminosities a diluted Planck function
is not a good fit to the model spectra (see
Figure~\ref{fig:modelfit}). Note that at the highest luminosities, the
color corrections of either work differ from those of Madej et al.\ by
about 5\%.
 
\section{Discussion/future directions}
\label{sec:discuss}

In this paper we model neutron star atmospheres at different times
during an XRB. For each model we specify an atmosphere radius,
composition, gravity, and equilibrium luminosity; guess the
atmosphere structure and radiation field; and then evolve the
atmosphere from our initial guess to the (quasi-)equilibrium by
solving the time-dependent equations of radiation transfer and
material energy along with the equation of hydrostatic balance
(Section~\ref{sec:equations}). This evolution proceeds in three steps,
as described in Section~\ref{sec:method}: 1) We evolve the atmosphere
to radiative equilibrium using the equations of radiation transfer and
material energy. We solve these equations in a coupled manner, using
an implicit Monte Carlo scheme that propagates photons in three
dimensions; the photons interact with the atmosphere material via
absorption/emission and scattering processes
(Section~\ref{sec:physics} and the Appendix). 2) We adjust the
material density profile using the equation of hydrostatic balance; to
solve this equation we include contributions from both gas and
radiation pressure. We then use step 1 to evolve the atmosphere to
radiative equilibrium for this new hydrostatic state. 3) We repeat
step 2 until a steady state is reached. If the luminosity from this
steady state is not the desired value, we adjust the temperature at
the base of the atmosphere and start the entire process over
again. The validity of our piecewise method (in particular, the
assumption that the atmosphere actually reaches a steady-state
structure during an XRB) is discussed in Sections~\ref{sub:radtrans}
and \ref{sub:hydro}.

For accuracy we use absorption opacities from the OPLIB database; the
additional bound-free and free-free transitions contained in the
database lead to differences between our results and those of
\citet{spw11,spw12} only at low luminosities and high metallicites
(see Section~\ref{sec:results}). We also use Compton scattering and
include the stimulated scattering contribution. Though stimulated
scattering is not typically included in Monte Carlo calculations
(because of the complexity it adds to the sampling algorithm for the
scattered photon; but see \citealt{wienke12}), we find its inclusion
necessary to obtain results that compare well with those of
\citet{spw12} (Sections~\ref{sub:opacity} and
\ref{sec:results}). Additionally, we consider general relativistic
effects within our model. In thin atmospheres ($r_{\rm surf} \alt
10^5$~cm), these effects can be added after calculation, as is done
by, e.g., \citet{madej04} and Suleimanov et al. However, in the
hottest XRBs, due to strong radiation pressure the atmosphere expands
to many times its pre-burst size \citep[e.g.,][]{gallowayetal2008},
such that the change in gravitational field across the atmosphere is
significant; in these atmospheres ($r_{\rm surf} \agt 10^6$~cm),
general relativity must be included both for accuracy and for
hydrostatic stability (Section~\ref{sub:gravity}).

We assume local thermodynamic equilibrium in the atmosphere, such that
the material radiates as a Planck function according to Kirchoff's law
[Equation~(\ref{eq:kirchoff})]. The outgoing radiation spectrum from
our models is non-Planckian, however, due to scattering and
frequency-dependent absorption of the radiation field before it
escapes the atmosphere (Section~\ref{sub:radmat}). At high
temperatures, absorption is weak and the majority of the photons come
from deep within the atmosphere; the color temperature of the outgoing
spectrum is much larger than the effective temperature (i.e., $f_{\rm
  c,proj} \gg 1$; Section~\ref{sec:results}).

Our time-dependent, stochastic (Monte Carlo) approach to modeling XRB
atmospheres differs from the time-independent, deterministic
approaches of \citet{madej04}, \citet{spw12}, and other groups, but
our goals are the same: to find the equilibrium solution for a given
set of atmosphere parameters. Our equilibrium spectra have the same
qualitative shape as those of Suleimanov et al., and differ generally
by less than 10\%; the color correction factors compare even more
closely (see Section~\ref{sec:results}). The fact that our results
compare so well with those of Suleimanov et al.\ despite the different
approaches, gives us more confidence in the validity of the methods
and solutions of our group and theirs. On the other hand, our results
do not compare well with those of Madej et al., which supports the
conclusion of Suleimanov et al.\ that the former group's results were
in error.

Note that even though our models are time dependent, they do not
represent the evolution of an atmosphere during a burst because we fix
the temperature at the base of the atmosphere, and because we change
the density profile in discrete jumps using a hydrostatic equation. To
model the burst evolution we would need to use time-dependent
temperature sources based on XRB energy generation models
\citep[e.g.,][]{woosleyetal2004,malone11,malone14}, as well as
solve the hydrodynamic equations together with our time-dependent
radiation transfer and material energy equations
(Section~\ref{sub:hydro}). We will consider these changes in future
work.

We have attempted to include the relevant pieces of physics in our
atmosphere models, but there is much we have left out for simplicity
(see Section~\ref{sec:physics}). For example, the composition of the
atmosphere in our models is a free parameter and is uniform in
space. In a real, X-ray-bursting neutron star atmosphere, the
composition varies with radius, due to a combination of different
processes: e.g., accretion, compositional mixing, and mass loss
(Section~\ref{sub:mixing}). Not having an accurate model for the
atmosphere composition introduces a large amount of uncertainty into
our calculation results, since as we saw in Section~\ref{sec:results},
the composition strongly affects both the shape of the outgoing
spectrum and the value of the color correction factor. While we
compensate for this uncertainty by generating a variety of models with
different compositions \citep[see also][]{spw11,spw12,nattila15}, this
is not ideal since current model fits to observations can not
distinguish between the possible compositions
\citep[e.g.,][]{zamfir12}. In future work, we hope to implement into
our atmosphere code a spatially varying composition that is motivated
by detailed physics calculations of the convection zone
\citep[e.g.,][]{woosleyetal2004,malone14}. As another example, the
outer layers of our atmosphere are not in thermal equilibrium with the
local radiation field. In the future we plan to implement opacities
that account for that fact, at least crudely
(Section~\ref{sub:plasma}).

In this paper we have attempted to reduce uncertainties in XRB models,
but there are many other uncertainties associated with XRBs that we
have not addressed or have addressed only partially (see
Section~\ref{sec:intro}). One uncertainty that we would like to
address more fully in the future is how the apparent expansion and
contraction of the atmosphere, characterized by the observed blackbody
radius $R_{\rm bb} = \left(L_\infty/4\pi \sigma_{\rm SB}T_{\rm
  bb}^4\right)^{1/2}$ and color temperature $T_{\rm bb}$ (see
\citealt{lewin93} for a review), correlates with the actual expansion
and contraction. During a PRE burst, $R_{\rm bb}$ grows to ${\rm
  a~few}\times 10$~km, then decays to a minimum value $\sim 10$~km,
then in some cases grows by ${\rm a~few}\times 10\%$, before finally
leveling off \citep[e.g., 4U~1724--307;
  see][]{gallowayetal2008}. Presumably the growth and decay of $R_{\rm
  bb}$ is correlated with the expansion and contraction of the
photosphere; however, this is complicated by the fact that $R_{\rm
  bb}$ also depends on the color correction factor, since the spectrum
is not a blackbody. In particular, since $R_{\rm bb} \propto f_{\rm
  c,proj}^{-2}$, if during contraction $f_{\rm c,proj}^2$ decays
faster than the photosphere radius does, the blackbody radius will
actually grow. This could mean that the time when $R_{\rm bb}$ reaches
a minimum, the so-called ``touchdown'' point, is not the same as the
time when the photosphere has returned to its pre-burst radius
\citep{steiner10,spw11,steiner13}. Even if the touchdown point
corresponds to a fully contracted photosphere, it is difficult to
derive constraints on the neutron star mass and radius from
observations only, without a detailed model of the atmosphere behavior
at touchdown. For example, some XRB groups
\citep[e.g.,][]{lewin93,ozel09} argue that the luminosity at touchdown
is given by the Eddington luminosity $L_{\rm Edd}$
[Equation~(\ref{eq:LEdd})], since above this luminosity the atmosphere
will be hydrodynamically unstable and therefore highly
extended. However, one can not use the Thomson approximation
$\kappa_{\rm Th}$ for the opacity in $L_{\rm Edd}$, since for neutron
stars near the Eddington limit the atmosphere will be hot and the true
opacity $\kappa_F^{\rm tot} \ll \kappa_{\rm Th}$
[Equation~(\ref{eq:kpapprox})]. Instead, for each atmosphere under
consideration, one must self-consistently find $L_{\rm Edd}$ at the
point where the atmosphere is on the edge of stability. We are
currently using our XRB code to model extended atmospheres like those
described by \citet{paczynski86} (see Section~\ref{sub:initial}), with
the goal of obtaining spectra and color correction factors at
luminosity ratios all the way up to the value at touchdown
\citep[which could potentially be as large as $1+z_{\rm base}$;
  cf.][]{lewin93,spw11}. Whether this can be done with a hydrostatic
code, or whether the full hydrodynamic equations must be solved,
remains to be seen.
 
\acknowledgements

This research was carried out in part under the auspices of the
National Nuclear Security Administration of the U.S.\ Department of
Energy at Los Alamos National Laboratory and supported by Contract
Nos.\ DE-AC52-06NA25396 and DE-FG02-87ER40317. Some of the results in
this paper were obtained using the high-performance computing system
at the Institute for Advanced Computational Science at Stony Brook.
 
\appendix

\section{Monte Carlo method for Compton scattering}
\label{sec:compton}

Here we describe our method for implementing Compton scattering within
the Monte Carlo framework, first when stimulated scattering is ignored
(Section~\ref{sub:nostim}), and then when it is included
(Section~\ref{sub:stimulated}). Our method is similar to that
of \citet{canfieldetal1987}, but we sample from a Maxwell-Boltzmann
electron distribution rather than a relativistic one. We refer
to \citet{kahn56,everett83,kalos86} for details on sampling from
Maxwell-Boltzmann and Klein-Nishina distributions.

\subsection{Ignoring stimulated scattering}
\label{sub:nostim}

To determine the probability of a scattering event occurring for a
specific Monte Carlo particle in the simulation, our code needs to
know the value of the scattering opacity $\kappa_\nu^{\rm
sc}$. Ignoring stimulated scattering, Equation~(\ref{eq:kappascat2})
becomes
\be
\kappa_\nu^{\rm sc} = \frac{Y_e}{m_u} \int_0^\infty 4\pi p^2 dp \, f(p) \int_{-1}^1 d\zeta \frac{1-\beta\zeta}{2} \sigma_{\rm KN}(\nu_0)
\label{eq:kappanostim}
\ee
with
\bal
\label{eq:KNnostim}
\sigma_{\rm KN}(\nu_0) = {}& \int_0^{2\pi} \frac{d\zeta_\perp}{2\pi} \int_0^{2\pi} \frac{d\eta_{0,\perp}}{2\pi} \int_{\nu_0/(1+2x_0)}^{\nu_0} d\nu'_0 \sigma_{\rm KN}(\nu_0 \rightarrow \nu'_0) \\
 = {}& \frac{3\sigma_{\rm Th}}{4x_0^2}\left[2 + \frac{x_0^2(1+x_0)}{(1+2x_0)^2} + \frac{x_0^2-2x_0-2}{2x_0}\ln(1+2x_0)\right]
\eal
\citep[e.g.,][]{rybickilightman1986}. Note that $\nu_0$, $\nu$,
$\zeta$, $\gamma$, $\beta$, and $p$ do not depend on $\zeta_\perp$ and
$\eta_{0,\perp}$ [Equations~(\ref{eq:p})--(\ref{eq:nu0})], such that
$\zeta_\perp$ and $\eta_{0,\perp}$ could be integrated out of
Equation~(\ref{eq:KNnostim}); however, we keep these latter variables
in the equation to demonstrate how our Monte Carlo sampling algorithm
works. We can express the integrand on the right-hand side of
Equation~(\ref{eq:kappanostim}) as a joint probability density
function
\be
P(\zeta,p) = P(\zeta|p)P(p)
\ee
times a rejection factor $R$ times a constant $C$:
\be
P(p) = 4\pi p^2 f(p)
\label{eq:Pp}
\ee
with
\be
\int_0^\infty P(p) \, dp = 1 \,,
\ee
\be
P(\zeta|p) = \frac{1}{2}(1-\beta\zeta)
\label{eq:Pz}
\ee
with
\be
\int_{-1}^1 d\zeta \, P(\zeta|p) = 1 \,,
\ee
\be
R = \frac{\sigma_{\rm KN}(\nu_0)}{\sigma_{\rm Th}} \le 1 \,,
\ee
and therefore
\be
C = \sigma_{\rm Th} \,.
\ee

We find the value of the integral in Equation~(\ref{eq:kappanostim})
through rejection sampling: We first sample from $P(p)$, a
Maxwell-Boltzmann distribution
\citep[e.g.,][]{everett83,kalos86}. Given $p$, we then sample from
$P(\zeta|p)$, by generating a random variable $\xi_1 \in [0,1]$ and
solving for $\zeta$ using
\be
\int_{-1}^\zeta d\bar{\zeta} \, P(\bar{\zeta}|p) = \frac{1}{2}\left[\zeta+1-\frac{\beta}{2}\left(\zeta^2-1\right)\right] = \xi_1 \,.
\ee
Finally, we generate another random variable $\xi_2 \in [0,1]$, and if
$\xi_2 > R$ we reject our $\{p,\zeta\}$ sample set. This process is
repeated many times, and we record the fraction of the total sample
sets that are accepted, $f_{\rm accept}$. After sufficient iterations
such that $f_{\rm accept}$ is converged ($10^4$ iterations or more),
the integral is given by $C \times f_{\rm accept}$, and
$\kappa_\nu^{\rm sc}$ is $Y_e/m_u$ times that. A table of
$\kappa_\nu^{\rm sc}$ values is calculated for a range of $\nu$ and
$T$, analogous to the table of absorption opacities. The table
generation is done only once, at the beginning of the simulation. Note
that we could instead use the scattering opacities from the OPLIB
database, also evaluated using
Equation~(\ref{eq:kappanostim}). However, we can not use the OPLIB
database for stimulated scattering (Section~\ref{sub:stimulated}), so
we use the above method for consistency.

When a scattering event occurs, the code also needs to know the
properties of the Monte Carlo particle after scattering, so that it
can determine the momentum and energy transferred between the particle
and the atmosphere material and can further transport the particle. We
first sample from $P(p)$ and $P(\zeta|p)$; then if $\xi_2 > R$, reject
the $\{p,\zeta\}$ set and resample until $\xi_2 \le R$. Note that
unlike above, we only do this procedure once (or rather, until the
first time $\xi_2 \le R$) for each scattering event. Then the
Klein-Nishina distribution is sampled \citep{kahn56,everett83} to find
$\nu'_0$ and $\eta_0$, and $\zeta_\perp$ and $\eta_{0,\perp}$ are each
sampled uniformly in $[0,2\pi]$. The photon direction after scattering
$\hat{\Omega}'$ is found through Lorentz transformations and
three-dimensional rotations: We first find $\hat{n}$ from
$\hat{\Omega}$, rotating $\hat{\Omega}$ by an angle $\arccos(\zeta)$
about the vector $\hat{\Omega}_{\perp1}$, where
$\hat{\Omega}_{\perp1}$ is some direction orthogonal to
$\hat{\Omega}$, to get $\hat{n}_1$; and then rotating $\hat{n}_1$ by
an angle $\zeta_\perp$ about the vector $\hat{\Omega}$. From
Rodrigues' rotation formula we have
\be
\hat{n}_1 = \zeta\hat{\Omega} + \sqrt{1-\zeta^2}\hat{\Omega}_{\perp2}
\ee
and
\be
\hat{n} = \zeta\hat{\Omega} + \sqrt{1-\zeta^2}\sin\zeta_\perp\hat{\Omega}_{\perp1} + \sqrt{1-\zeta^2}\cos\zeta_\perp\hat{\Omega}_{\perp2} \,,
\ee
where $\hat{\Omega}_{\perp2}$ is orthogonal to both
$\hat{\Omega}_{\perp1}$ and $\hat{\Omega}$. For simplicity we choose
$\hat{\Omega}_{\perp1}$ to be in the $x$-$y$ plane, such that in
Cartesian coordinates
\be
\hat{n} = \left(
\begin{array}{c}
\zeta\Omega_x + \frac{\sqrt{1-\zeta^2}}{\sqrt{1-\Omega_z^2}}\left(-\sin\zeta_\perp\Omega_y + \cos\zeta_\perp\Omega_x\Omega_z\right) \\
\zeta\Omega_y + \frac{\sqrt{1-\zeta^2}}{\sqrt{1-\Omega_z^2}}\left(\sin\zeta_\perp\Omega_x + \cos\zeta_\perp\Omega_y\Omega_z\right) \\
\zeta\Omega_z - \sqrt{1-\zeta^2}\sqrt{1-\Omega_z^2}\cos\zeta_\perp
\end{array}
\right)
\ee
(unless $\hat{\Omega} \equiv \pm\hat{z}$, in which case we choose
$\hat{n} = \pm\zeta\hat{z} + \sqrt{1-\zeta^2}\sin\zeta_\perp\hat{y}
+ \sqrt{1-\zeta^2}\cos\zeta_\perp\hat{x}$). Note that the ambiguity in
the value of $\sin(\arccos(\zeta)) = \pm\sqrt{1-\zeta^2}$ and the
choice of $\hat{\Omega}_{\perp1}$, is accounted for by the fact that
$\zeta_\perp$ is a random number between $0$ and $2\pi$; we can define
$\arccos(\zeta)$ over any domain and choose any orthogonal vector
$\hat{\Omega}_{\perp1}$ and still cover all possible directions for
$\hat{n}$. After finding $\hat{n}$ we find $\hat{\Omega}_0$ (the
photon direction pre-scattering and in the electron rest frame), using
the Lorentz angle transformations \citep[e.g.,][]{rybickilightman1986}
\be
\zeta_0 = \frac{\zeta-\beta}{1-\beta\zeta} = \frac{\nu}{\nu_0}\gamma(\zeta-\beta)
\ee
and
\be
\zeta_{0,\perp} = \frac{\zeta_\perp}{\gamma(1-\beta\zeta)} = \frac{\nu}{\nu_0}\zeta_\perp \,;
\ee
combined, these give
\be
\hat{\Omega}_0 = \frac{\nu}{\nu_0}\left\{\hat{\Omega} + [(\gamma-1)\zeta-\gamma\beta]\hat{n} \right\} \,.
\ee
We find $\hat{\Omega}'_0$ (the photon direction post-scattering and in
the electron rest frame) using Rodrigues' rotation formula in the same
manner as described above to get
\be
\hat{\Omega}'_0 = \left(
\begin{array}{c}
\eta_0\Omega_{0,x} + \frac{\sqrt{1-\eta_0^2}}{\sqrt{1-\Omega_{0,z}^2}}\left(-\sin\eta_{0,\perp}\Omega_{0,y} + \cos\eta_{0,\perp}\Omega_{0,x}\Omega_{0,z}\right) \\
\eta_0\Omega_{0,y} + \frac{\sqrt{1-\eta_0^2}}{\sqrt{1-\Omega_{0,z}^2}}\left(\sin\eta_{0,\perp}\Omega_{0,x} + \cos\eta_{0,\perp}\Omega_{0,y}\Omega_{0,z}\right) \\
\eta_0\Omega_{0,z} - \sqrt{1-\eta_0^2}\sqrt{1-\Omega_{0,z}^2}\cos\eta_{0,\perp}
\end{array}
\right) \,.
\ee
Finally, we find $\hat{\Omega}'$ using Lorentz angle transformations
in the same manner as above (but transforming back to the neutron star
frame) to get
\be
\hat{\Omega}' = \frac{\nu'_0}{\nu'}\left\{\hat{\Omega}_0' + [(\gamma-1)\zeta'_0+\gamma\beta]\hat{n} \right\} \,,
\ee
where
\be
\nu' = \gamma \nu_0'(1+\beta\zeta'_0)
\ee
and
\be
\zeta'_0 = \hat{\Omega}'_0 \cdot \hat{n} \,.
\ee

\subsection{Including stimulated scattering}
\label{sub:stimulated}

To find $\kappa_\nu^{\rm sc}$ when stimulated scattering is included,
we must solve Equation~(\ref{eq:kappascat2}) as written. Here we can
not integrate $\nu'_0$ out of the scattering equation (since the
integral with respect to $\nu'_0$ is not analytic), so we express the
integrand on the right-hand side of the equation as
\be
P(\nu'_0,\eta_{0,\perp},\zeta_\perp,\zeta,p) = P(\nu'_0|\eta_{0,\perp},\zeta_\perp,\zeta,p)P(\eta_{0,\perp}|\zeta_\perp,\zeta,p)P(\zeta_\perp|\zeta,p)P(\zeta|p)P(p)
\ee
times a rejection factor $R$ times a constant $C$; $P(p)$ and
$P(\zeta|p)$ are as in Equations~(\ref{eq:Pp}) and (\ref{eq:Pz}),
\be
P(\zeta_\perp|\zeta,p) = \frac{1}{2\pi}
\ee
with
\be
\int_0^{2\pi} P(\zeta_\perp|\zeta,p) \, d\zeta_\perp = 1 \,,
\ee
\be
P(\eta_{0,\perp}|\zeta_\perp,\zeta,p) = \frac{1}{2\pi}
\ee
with
\be
\int_0^{2\pi} P(\eta_{0,\perp}|\zeta_\perp,\zeta,p) \, d\eta_{0,\perp} = 1 \,,
\ee
\be
P(\nu'_0,\eta_{0,\perp},\zeta_\perp,\zeta,p) = \frac{\sigma_{\rm KN}(\nu_0 \rightarrow \nu'_0)}{\sigma_{\rm KN}(\nu_0)}
\ee
with
\be
\int_{\nu_0/(1+2x_0)}^{\nu_0} d\nu'_0 \, P(\nu'_0|\eta_{0,\perp},\zeta_\perp,\zeta,p) = 1 \,,
\ee
\be
R = \frac{1+c^2I_{\nu'}(\hat{\Omega}')/2h{\nu'}^3}{1+[\exp(Ah\nu/k_{\rm B}T)-1]^{-1}} \frac{\sigma_{\rm KN}(\nu_0)}{\sigma_{\rm Th}} \alt 1 \,,
\label{eq:Rstim}
\ee
and therefore
\be
C = \left[1+\frac{1}{\exp(Ah\nu/k_{\rm B}T)-1}\right]\sigma_{\rm Th} \,.
\ee
Here $A \le 1$ is a parameter chosen by the user (see below). To
reduce the noise on $I_\nu(\hat{\Omega})$ we average this quantity
over 1000 time steps (cf.\ Section~\ref{sub:MC}); we therefore also
only recalculate the $\kappa_\nu^{\rm sc}$ table every 1000 time
steps, to save calculation time. We sample the various probability
density functions above as in Section~\ref{sub:nostim}
[$P(\nu'_0|\eta_{0,\perp},\zeta_\perp,\zeta,p)$ is the Klein-Nishina
distribution]; then repeat the sampling many times until we obtain a
converged value of $f_{\rm accept}$, after which we set
$\kappa_\nu^{\rm sc} = (Y_e/m_u)Cf_{\rm accept}$. For a single
scattering event, we sample the various functions once; then if $\xi >
R$ reject the sample set and resample until $\xi \le R$.

The approximation in our treatment of stimulated scattering is that
$R$ in Equation~(\ref{eq:Rstim}) never exceeds unity. The inequality
\be
\sigma_{\rm KN}(x) \le \sigma_{\rm Th}
\ee
is always true, but
\be
\frac{c^2I_{\nu'}(\hat{\Omega}')}{2h{\nu'}^3} \le \frac{1}{\exp(Ah\nu/k_{\rm B}T)-1}
\ee
is an empirical relation (chosen because $c^2I_\nu/2h\nu^3 =
[\exp(h\nu/k_{\rm B}T)-1]^{-1}$ for a Planckian radiation field) that
is only absolutely true in the limit $A \rightarrow 0$. In practical
terms, the parameter $A$ should not be too small or the number of
rejections will be very large and the Monte Carlo integration will
take a long time; but it should not be too large or the error in the
result will be large. For our XRB atmosphere calculations, we chose $A
= 0.2$ for reasonably fast integration time and a rejection factor
Equation~(\ref{eq:Rstim}) that exceeds unity less than 1 time in
$10^5$.

\bibliography{master}

\end{document}